\documentclass{ws-mpla}
\usepackage[super]{cite}
\usepackage{graphicx}

\usepackage{dsfont}
\usepackage{tikz}

\begin{document}

	\markboth{M. F. Araujo de Resende}{Finite-group gauge theories on lattices as Hamiltonian systems with constraints}

	%%%%%%%%%%%%%%%%%%%%% Publisher's Area please ignore %%%%%%%%%%%%%%
	\catchline{}{}{}{}{}
	%%%%%%%%%%%%%%%%%%%%%%%%%%%%%%%%%%%%%%%%%%%%%%%%%%%%%%%%%%%%%%%%%%%

	\title{Finite-group gauge theories on lattices \\ as Hamiltonian systems with constraints}

	\author{M. F. Araujo de Resende}

	\address{Instituto de Física, Universidade de São Paulo, 05508-090 São Paulo SP, Brasil \\ resende@if.usp.br}

	\maketitle

	\pub{Received (11 September 2022)}{Revised (Day Month Year)}

	\begin{abstract}
		In this work, we present a brief but insightful overview of the gauge theories, which are defined on $ n $-dimensional lattices by using finite gauge groups, in order to show how they can be interpreted as a Hamiltonian system with constraints, analogous to what happens with the classical (continuous) gauge (field) theories. As this interpretation is not usually explored in the literature that discusses/introduces the concept of lattice gauge theory, but some recent works have been exploring Hamiltonian models in order to support some kind of quantum computation, we use this interpretation to, for example, present a brief geometric view of one class of these models: the Kitaev Quantum Double Models.

	\end{abstract}

	\ccode{PACS Nos.: 11.10.Ef, 11.15.Ha.}

	\section{\label{intro} Introduction}
	
		One of the ingredients that characterizes a classical system as a gauge theory is its restriction on some submanifold $ \mathcal{M} _{n} \subset \mathcal{M} _{n+k} $ with $ n $ dimensions [\refcite{mf-gauge}]. And when we describe this system by using a Hamiltonian formulation, this restriction takes its form through a Hamiltonian function [\refcite{gitman}]
		\begin{equation}
			H_{\mathrm{T}} \left( z \right) = H \left( z \right) + \lambda ^{j} \Phi _{j} \left( z \right) \ , \label{pre-H-total}
		\end{equation}
		where $ \mathcal{S} _{\Phi } = \bigl\{ \Phi _{j} : T^{\ast } \mathcal{M} _{n+k} \rightarrow \mathbb{R} \bigr\} $ is a set of independent differentiable functions, which are responsible for defining the phase subspace $ T^{\ast } \mathcal{M} _{n} \subset T^{\ast } \mathcal{M} _{n+k} $ of this system by taking $ \Phi _{j} \left( z \right) = 0 $, and $ j = 1 , \ldots , k $ [\refcite{deriglazov}]. Note that, as $ T^{\ast } \mathcal{M} _{n+k} $ can be parameterized by more than one atlas/coordinate system, the functions that describe this classical system can be expressed in a non-unique way. And one of the consequences is that, due to the differentiable decomposition [\refcite{man-gr}]
		\begin{equation}
			T^{\ast } _{\mathsf{q}} \mathcal{M} _{n+k} = T^{\ast } _{\mathsf{q}} \mathcal{M} _{n} \oplus \left( T^{\ast } _{\mathsf{q}} \mathcal{M} _{n} \right) ^{\perp } \label{decomposition}
		\end{equation}
		that these constraints $ \Phi _{j} \left( z \right) = 0 $ allow us to do, where $ \mathsf{q} $ is a point that belongs to $ \mathcal{M} _{n} $, there is another parametrization
		\begin{equation*}
			\kappa = \left( \mathcal{Q} , \mathcal{P} \right) = \left( \right. \underbrace{q,Q} _{\mathcal{Q}} \hspace*{0.04cm} , \hspace*{0.04cm} \underbrace{p,P} _{\mathcal{P}} \left. \right)
		\end{equation*}
		of $ T^{\ast } \mathcal{M} _{n+k} $ that, for example, allows us to rewrite all the physical functions of this same system by using only the intrinsic parameters $ \omega = \left( q , p \right) $ of the cotangent bundle $ T^{\ast } \mathcal{M} _{n} $ [\refcite{mf-gauge}]. In this case, these new parametrization let us rewrite (\ref{pre-H-total}) as\footnote{Here, we are taking advantage of the fact that, by rearranging the entries of $ \left( \mathcal{Q} , \mathcal{P} \right) $, we can rewrite it as $ \left( \omega , \Omega \right) = \left( q , p , Q , P \right) $, where $ \Omega = \left( Q , P \right) $ parameterizes $ \left( T^{\ast } \mathcal{M} _{n} \right) ^{\perp } $ intrinsically [\refcite{man-gr}].} [\refcite{gitman,teitel}]
		\begin{equation}
			H^{\prime } _{T} \left( \kappa \right) = H_{\mathrm{ph}} \left( \omega \right) + \lambda _{P} P + \mathcal{O} \bigl( \dot{P} , P^{2} \bigr) \label{hamiltonian-reexpress}
		\end{equation}
		
		However, it is worth to remember that these are not the only ingredients necessary to characterize a classical gauge theory as a Hamiltonian system with constraints: it is also necessary that, at least, a part of the functions in the set $ \mathcal{S} _{\Phi } $ be \emph{first-class} [\refcite{dirac}] because the Lagrange multipliers ($ \lambda _{I} $), which implement these first-class constraints ($ \Phi _{I} $) to the Hamiltonian function (\ref{pre-H-total}), can never be solved unequivocally [\refcite{gitman}]. In this way, by noting that all the parametrizations of a manifold are related (one to the other) through \emph{diffeomorphisms} [\refcite{elon-variedades}], it is not difficult to conclude that, due to the bijection between the components of $ P $ and $ \Phi = \left( \Phi _{1} , \ldots , \Phi _{k} \right) $, the new constraints $ P=0 $ can also be divided between those that are of first- ($ P_{I} $) and second-class ($ P_{II} $). As a consequence, this new Hamiltonian function (\ref{hamiltonian-reexpress}) describes the same classical system through a set of new equations [\refcite{tyutin}]
		\begin{equation}
			\dot{\omega } = \left\{ \omega , H_{\mathrm{ph}} \right\} \ , \quad \dot{Q} _{I} = \lambda _{P_{I}} \ , \quad \dot{Q} _{II} = \mathcal{A} \left( \omega , Q \right) \ , \quad \textnormal{and} \quad P = 0 \label{eq-motion}
		\end{equation}
		that are very interesting. And why are these new equations so interesting? Because,
		\begin{itemize}
			\item by noting that $ \omega = \left( q , p \right) $ and $ \Omega = \left( Q , P \right) $ are the intrinsic parameters of $ T^{\ast } \mathcal{M} _{n} $ and $ \left( T^{\ast } \mathcal{M} _{n} \right) ^{\perp } $ respectively, and
			\item as $ \lambda _{P_{I}} $ designates the new non-univocal Lagrange multipliers that implement the new first-class constraints $ P_{I} = 0 $ in (\ref{hamiltonian-reexpress}),
		\end{itemize}
		this shows us that the endless choices that we can make for these multipliers and, consequently, for the \emph{gauge} $ Q = \left( Q_{I} , Q_{II} \right) $ and $ P = \left( P_{I} , P_{II} \right) $ never change the solution of the physical equations\footnote{Note that, just as $ \omega = \left( q , p \right) $ can be interpreted as a canonical pair of variables, so can $ \Omega _{I} = \left( Q_{I} , P_{I} \right) $ and $ \Omega _{II} = \left( Q_{II} , P_{II} \right) $. Therefore, $ \left( Q_{I} , Q_{II} \right) $ and $ \left( P_{I} , P_{II} \right) $ must be interpreted as the components of the canonical pair of variables $ \Omega = \left( Q , P \right) $.} [\refcite{tyutin}]
		\begin{equation}
			\dot{\omega } = \left\{ \omega , H_{\mathrm{ph}} \right\} \ . \label{eq-physics}
		\end{equation} 
		And by remembering, once again, that all the different parametrizations of a manifold are related by diffeomorphisms, it is exactly this freedom (which we have to fix any gauge $ \Omega = \left( Q , P \right) $) that ensures that infinitely many other gauges $ \Omega ^{\prime } = \left( Q^{\prime } , P^{\prime } \right) $ can also be chosen in all other parametrizations [\refcite{mf-gauge}].
		
		Nevertheless, once it is already well-known that this constrained interpretation extends to the continuous gauge field theories\footnote{That is, gauge theories where all their Hamiltonian functions are defined by using Hamiltonian densities.} [\refcite{gitman,teitel}], it is interesting to go in the \textquotedblleft opposite\textquotedblright \hspace*{0.01cm} direction in order to show, for instance, how this same interpretation fits with the physical systems that behave as gauge theories on \emph{lattices}: i.e., as theories whose gauge fields are attached to the edges of some spatial lattice $ \mathcal{L} _{n} $ with $ n $ dimensions [\refcite{rothe}]. And this is exactly what we will do throughout this review (i) by analysing a pure gauge theory on this $ \mathcal{L} _{n} $ and (ii) by assuming that this lattice is a kind of \textquotedblleft patchwork quilt\textquotedblright \hspace*{0.01cm} composed of not necessarily regular polyhedra that (locally) discretizes some (sub)manifold $ \mathcal{M} _{n} $. After all, in addition to the fact that there are not many references that go in this \textquotedblleft opposite\textquotedblright \hspace*{0.01cm} direction, this constrained interpretation is of great value due, for instance, to the recent interest in using Hamiltonian models, which describe finite-group gauge theories on spatial lattices, that try to support some kind of quantum computing [\refcite{bruno,zohar,pramod-non-abelian,cubitt,brower,lam,banuls,meurice,atas,gustafson,halimeh,lumia,carena,mf-annals}]. We will give an example of such Hamiltonian models in the penultimate Section of this manuscript. %%%%%% 89B
		
	\section{\label{lattice} Gauge theories on lattices}
	
		Roughly speaking, it is not wrong to say that lattice gauge theories emerged from the need to solve some problems that could not be solved by using the same formulation as continuous gauge field theories. And, among these problems, we can list those that were directly related to elementary particle physics (in particular, to quantum chromodynamics) that, for example, required a non-perturbative solution [\refcite{rothe}]. Thus, as a solution for these problems, these continuous gauge theories were replicated in a new scenario where this was possible: strictly speaking, to a scenario where, instead of considering that a physical system evolves on $ \mathcal{M} _{n} $, it was considered that this physical system evolves on the discretization $ \mathcal{L} _{n} $ of this manifold [\refcite{wilson-loops}].
		
		Among the adaptations that had to be made in this new scenario, probably the most basic was to assume that all matter could be represented by fields assigned only to the vertices of this lattice $ \mathcal{L} _{n} $. However, given the need to adapt the Lagrangian/Hamiltonian formulation to this new scenario and, therefore, identify the symmetry transformations performed by the new gauge fields, all these gauge fields ended up being exclusively assigned to the lattice edges [\refcite{fradkin}]. After all, as the set of constraints $ \Phi = 0 $ (that define $ T^{\ast } \mathcal{M} _{n} \subset T^{\ast } \mathcal{M} _{n+k} $ in the continuous gauge theories) generates symmetries that can be controlled by some (Lie) group [\refcite{gitman,teitel,castelani}], the replication of these continuous gauge theories in this new scenario forces, for instance, that these lattice gauge fields be elements of some group $ G $ since they need to be interpreted as \emph{parallel transporters} [\refcite{montvay,knowles}].
		
		\subsection{Gauge theories on lattices defined by using finite gauge groups}
		
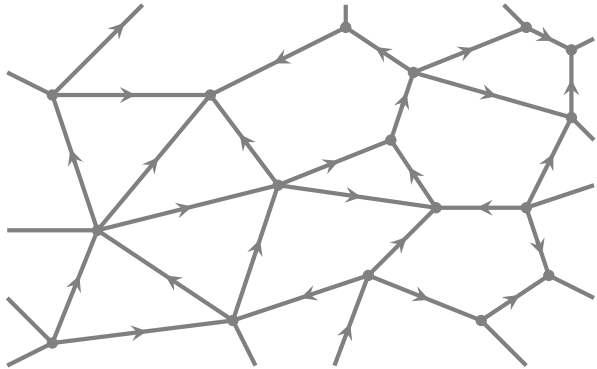
\begin{figure}[!t]
				\begin{center}
					\begin{tikzpicture}[
						scale=0.3,
						trans/.style={color=gray, ultra thick},
						function/.style={->, color=gray, ultra thick, >=stealth}
						]
						\draw[trans,color=gray] (0,0) -- (2,1) node[above=2pt,right=-1pt] {};
						\draw[trans,color=gray] (2,1) -- (0,3) node[above=2pt,right=-1pt] {};
						\draw[function,color=gray] (2,1) -- (6.2,1.525) node[above=2pt,right=-1pt] {};
						\draw[trans,color=gray] (6,1.5) -- (10,2) node[above=2pt,right=-1pt] {};
						\draw[function,color=gray] (2,1) -- (3.2,4) node[above=2pt,right=-1pt] {};
						\draw[trans,color=gray] (3,3.5) -- (4,6) node[above=2pt,right=-1pt] {};
						\draw[trans,color=gray] (7.2,3.867) -- (4,6) node[above=2pt,right=-1pt] {};
						\draw[function,color=gray] (10,2) -- (7,4) node[above=2pt,right=-1pt] {};
						\draw[trans,color=gray] (10,2) -- (11,0) node[above=2pt,right=-1pt] {}; 
						\draw[trans,color=gray] (13.2,3.067) -- (10,2) node[above=2pt,right=-1pt] {};
						\draw[function,color=gray] (16,4) -- (13,3) node[above=2pt,right=-1pt] {};
						\draw[function,color=gray] (10,2) -- (11.2,5.6) node[above=2pt,right=-1pt] {};
						\draw[trans,color=gray] (11,5) -- (12,8) node[above=2pt,right=-1pt] {};
						\draw[function,color=gray] (14.5,0) -- (15.2,1.867) node[above=2pt,right=-1pt] {};
						\draw[trans,color=gray] (15,1.333) -- (16,4) node[above=2pt,right=-1pt] {};
						\draw[function,color=gray] (16,4) -- (17.7,5.7) node[above=2pt,right=-1pt] {}; 
						\draw[trans,color=gray] (17.5,5.5) -- (19,7) node[above=2pt,right=-1pt] {}; 
						\draw[function,color=gray] (12,8) -- (15.7,7.471) node[above=2pt,right=-1pt] {};
						\draw[trans,color=gray] (15.5,7.5) -- (19,7) node[above=2pt,right=-1pt] {};
						\draw[trans,color=gray] (4,6) -- (0,6) node[above=2pt,right=-1pt] {};
						\draw[function,color=gray] (4,6) -- (8.2,7.05) node[above=2pt,right=-1pt] {};
						\draw[trans,color=gray] (8,7) -- (12,8) node[above=2pt,right=-1pt] {};
						\draw[function,color=gray] (4,6) -- (2.8,9.6) node[above=2pt,right=-1pt] {};
						\draw[trans,color=gray] (3,9) -- (2,12) node[above=2pt,right=-1pt] {};
						\draw[function,color=gray] (4,6) -- (6.7,9.24) node[above=2pt,right=-1pt] {};
						\draw[trans,color=gray] (6.5,9) -- (9,12) node[above=2pt,right=-1pt] {};
						\draw[function,color=gray] (12,8) -- (10.3,10.267) node[above=2pt,right=-1pt] {};
						\draw[trans,color=gray] (10.5,10) -- (9,12) node[above=2pt,right=-1pt] {};
						\draw[function,color=gray] (12,8) -- (14.7,9.08) node[above=2pt,right=-1pt] {};
						\draw[trans,color=gray] (14.5,9) -- (17,10) node[above=2pt,right=-1pt] {};
						\draw[function,color=gray] (16,4) -- (18.7,2.92) node[above=2pt,right=-1pt] {};
						\draw[trans,color=gray] (18.5,3) -- (21,2) node[above=2pt,right=-1pt] {};
						\draw[trans,color=gray] (21,2) -- (23,0) node[above=2pt,right=-1pt] {};
						\draw[function,color=gray] (21,2) -- (22.7,3.133) node[above=2pt,right=-1pt] {};
						\draw[trans,color=gray] (22.5,3) -- (24,4) node[above=2pt,right=-1pt] {};
						\draw[function,color=gray] (23,7) -- (23.7,4.9) node[above=2pt,right=-1pt] {};
						\draw[trans,color=gray] (23.5,5.5) -- (24,4) node[above=2pt,right=-1pt] {};
						\draw[function,color=gray] (23,7) -- (20.8,7) node[above=2pt,right=-1pt] {};
						\draw[trans,color=gray] (21,7) -- (19,7) node[above=2pt,right=-1pt] {};
						\draw[trans,color=gray] (24,4) -- (26,3) node[above=2pt,right=-1pt] {}; 
						\draw[trans,color=gray] (23,7) -- (26,8) node[above=2pt,right=-1pt] {}; 
						\draw[function,color=gray] (19,7) -- (17.8,8.8) node[above=2pt,right=-1pt] {}; 
						\draw[trans,color=gray] (18,8.5) -- (17,10) node[above=2pt,right=-1pt] {};
						\draw[function,color=gray] (17,10) -- (17.7,12.1) node[above=2pt,right=-1pt] {};
						\draw[trans,color=gray] (17.5,11.5) -- (18,13) node[above=2pt,right=-1pt] {};
						\draw[function,color=gray] (18,13) -- (16.3,14.133) node[above=2pt,right=-1pt] {};
						\draw[trans,color=gray] (16.5,14) -- (15,15) node[above=2pt,right=-1pt] {};
						\draw[function,color=gray] (15,15) -- (11.8,13.4) node[above=2pt,right=-1pt] {};
						\draw[trans,color=gray] (12,13.5) -- (9,12) node[above=2pt,right=-1pt] {};
						\draw[function,color=gray] (2,12) -- (5.7,12) node[above=2pt,right=-1pt] {};
						\draw[trans,color=gray] (5.5,12) -- (9,12) node[above=2pt,right=-1pt] {};
						\draw[function,color=gray] (2,12) -- (5.2,15.2) node[above=2pt,right=-1pt] {};
						\draw[trans,color=gray] (5,15) -- (6,16) node[above=2pt,right=-1pt] {};
						\draw[trans,color=gray] (15,15) -- (15,16) node[above=2pt,right=-1pt] {}; 
						\draw[function,color=gray] (18,13) -- (20.7,14.08) node[above=2pt,right=-1pt] {};
						\draw[trans,color=gray] (20.5,14) -- (23,15) node[above=2pt,right=-1pt] {};
						\draw[trans,color=gray] (23,15) -- (22,16) node[above=2pt,right=-1pt] {}; 
						\draw[function,color=gray] (23,15) -- (24.2,14.4) node[above=2pt,right=-1pt] {};
						\draw[trans,color=gray] (24,14.5) -- (25,14) node[above=2pt,right=-1pt] {};
						\draw[trans,color=gray] (25,14) -- (26,14.5) node[above=2pt,right=-1pt] {}; 
						\draw[function,color=gray] (23,7) -- (24.2,9.4) node[above=2pt,right=-1pt] {};
						\draw[trans,color=gray] (24,9) -- (25,11) node[above=2pt,right=-1pt] {};
						\draw[function,color=gray] (25,11) -- (25,12.7) node[above=2pt,right=-1pt] {}; 
						\draw[trans,color=gray] (25,12.5) -- (25,14) node[above=2pt,right=-1pt] {};
						\draw[function,color=gray] (18,13) -- (21.7,11.943) node[above=2pt,right=-1pt] {};
						\draw[trans,color=gray] (21.5,12) -- (25,11) node[above=2pt,right=-1pt] {};
						\draw[trans,color=gray] (25,11) -- (26,10) node[above=2pt,right=-1pt] {};
						\draw[trans,color=gray] (2,12) -- (0,13) node[above=2pt,right=-1pt] {};
						\draw[color=gray,fill=gray] (2,1) circle (1.5ex);
						\draw[color=gray,fill=gray] (4,6) circle (1.5ex);
						\draw[color=gray,fill=gray] (10,2) circle (1.5ex);
						\draw[color=gray,fill=gray] (16,4) circle (1.5ex); 
						\draw[color=gray,fill=gray] (12,8) circle (1.5ex);
						\draw[color=gray,fill=gray] (2,12) circle (1.5ex);
						\draw[color=gray,fill=gray] (23,7) circle (1.5ex);
						\draw[color=gray,fill=gray] (19,7) circle (1.5ex);
						\draw[color=gray,fill=gray] (17,10) circle (1.5ex); 
						\draw[color=gray,fill=gray] (18,13) circle (1.5ex);
						\draw[color=gray,fill=gray] (15,15) circle (1.5ex); 
						\draw[color=gray,fill=gray] (9,12) circle (1.5ex); 
						\draw[color=gray,fill=gray] (23,15) circle (1.5ex);
						\draw[color=gray,fill=gray] (25,14) circle (1.5ex); 
						\draw[color=gray,fill=gray] (25,11) circle (1.5ex); 
						\draw[color=gray,fill=gray] (24,4) circle (1.5ex);
						\draw[color=gray,fill=gray] (21,2) circle (1.5ex);
					\end{tikzpicture}
				\end{center}
				\caption{\protect\label{oriented-lattice} Example of a $ 2 $-dimensional lattice $ \mathcal{L} _{2} $ where all the edges are oriented. Note that, since each of the edges of this \textquotedblleft patchwork quilt\textquotedblright \hspace*{0.01cm} can be oriented in two ways, this example makes it clear that there is no rule for making these orientations. As will become clear in the following lines, these edge orientations are necessary not only so that the gauge fields can be interpreted as parallel transporters, but also so that, among other things, the lattice gauge transformations can be well-defined. In any case, it is worth noting that, although the spatial lattices that define the lattice gauge theories are usually interpreted as regular hypercubic lattices, we are using this example to reinforce that we will not restrict considerations to only regular hypercubic lattices: in this paper, we will only consider that $ \mathcal{L} _{n} $ is a spatial lattice that discretizes an $ n $-dimensional manifold.}
			\end{figure}

			A simple but important thing we must notice here is that, due to this interpretation of the lattice gauge fields as parallel transporters, $ \mathcal{L} _{n} $ should be an oriented lattice: i.e., all their edges must be oriented as illustrated, for instance, in Figure \ref{oriented-lattice}.
			And the best way for us to begin to understand why this orientation needs to be done is to recognize, first, that one of the main consequences of this lattice approach is that, when $ G $ is a finite group, we can obtain the Hamiltonian formulation of these lattice gauge theories through a \emph{partition function} [\refcite{rothe,salinas}]
			\begin{equation}
				Z = \sum _{\left\{ \mathfrak{g} \right\}} e^{-\beta S \left( \mathfrak{g} \right) } \ , \label{pre-part-function}
			\end{equation}
			where $ S $ is the action that describes this system and $ \beta $ is a real constant. And as this action is written as a function of the lattice gauge fields $ \mathfrak{g} \in G $, the sum variable $ \left\{ \mathfrak{g} \right\} $ in (\ref{pre-part-function}) means that all the possible configurations of these fields are being computed in $ Z $.
			
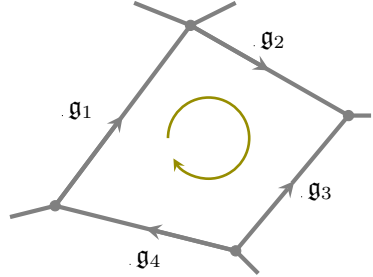
\begin{figure}[!t]
				\begin{center}
					\begin{tikzpicture}[
						scale=0.3,
						equation/.style={thin},
						trans/.style={color=gray, ultra thick},
						function/.style={->, color=gray, ultra thick, >=stealth}
						]
						\draw [color=olive,thin,->,line width=1.0pt,>=stealth] (21,3) arc[x radius=1.8cm, y radius =1.8cm, start angle=180, end angle=-150];
						\draw[function,color=gray] (24,-2) -- (20,-1.0) node[above=2pt,right=-1pt] {};
						\draw[function,color=gray] (16,0) -- (19.0,4) node[above=2pt,right=-1pt] {};
						\draw[function,color=gray]  (24,-2) -- (26.6,1.1) node[above=2pt,right=-1pt] {};
						\draw[function,color=gray] (22,8) -- (25.5,6) node[above=2pt,right=-1pt] {};
						\draw[trans,color=gray] (16,0) -- (14,-0.5) node[above=2pt,right=-1pt] {};
						\draw[trans,color=gray] (24,-2) -- (25,-3) node[above=2pt,right=-1pt] {};
						\draw[trans,color=gray] (22,8) -- (24,9) node[above=2pt,right=-1pt] {};
						\draw[trans,color=gray] (22,8) -- (19.5,9) node[above=2pt,right=-1pt] {};
						\draw[trans,color=gray] (29,4) -- (30,4) node[above=2pt,right=-1pt] {};	
						\draw[trans,color=gray] (16,0) -- (24,-2) node[above=2pt,right=-1pt] {};
						\draw[trans,color=gray] (16,0) -- (22,8) node[above=2pt,right=-1pt] {};
						\draw[trans,color=gray] (22,8) -- (29,4) node[above=2pt,right=-1pt] {};
						\draw[trans,color=gray] (26.5,1.0) -- (29.0,4) node[above=2pt,right=-1pt] {};
						\draw[equation] (16.3,4.0) -- (16.3,4.0) node[above=2pt,right=-1pt] {$ \mathfrak{g} _{1} $};
						\draw[equation] (24.8,7.1) -- (24.8,7.1) node[above=2pt,right=-1pt] {$ \mathfrak{g} _{2} $};
						\draw[equation] (26.9,0.4) -- (26.9,0.4) node[above=2pt,right=-1pt] {$ \mathfrak{g} _{3} $};
						\draw[equation] (19.3,-2.6) -- (19.3,-2.6) node[above=2pt,right=-1pt] {$ \mathfrak{g} _{4} $};
						\draw[color=gray,fill=gray] (16,0) circle (1.5ex);
						\draw[color=gray,fill=gray] (24,-2) circle (1.5ex);
						\draw[color=gray,fill=gray] (22,8) circle (1.5ex);
						\draw[color=gray,fill=gray] (29,4) circle (1.5ex);
					\end{tikzpicture}
				\end{center}
				\caption{\protect\label{clockwise-holonomy} Here, we see a lattice face $ f $ that is delimited by four edges, where it is possible to identify a gauge field $ \mathfrak{g} _{\ell } $ on each of them. Note that, if we calculate their face holonomies by using the clockwise direction (highlighted in olive colour), the four possible results are \\ \hspace*{2.0cm} $ U^{\left( 1a \right) } _{f} = \mathfrak{g} _{1} \cdot \mathfrak{g} _{2} \cdot \left( \mathfrak{g} _{3} \right) ^{-1} \cdot \mathfrak{g} _{4} $ \ , \ \ $ U^{\left( 2a \right) } _{f} = \mathfrak{g} _{2} \cdot \left( \mathfrak{g} _{3} \right) ^{-1} \cdot \mathfrak{g} _{4} \cdot \mathfrak{g} _{1} $ \ , \\ \hspace*{2.0cm} $ U^{\left( 3a \right) } _{f} = \left( \mathfrak{g} _{3} \right) ^{-1} \cdot \mathfrak{g} _{4} \cdot \mathfrak{g} _{1} \cdot \mathfrak{g} _{2} $ \ \ and \ \ $ U^{\left( 4a \right) } _{f} = \mathfrak{g} _{4} \cdot \mathfrak{g} _{1} \cdot \mathfrak{g} _{2} \cdot \left( \mathfrak{g} _{3} \right) ^{-1} $ \ . \\ This calculation was performed by noting that, when the orientation of the edge that contains $ \mathfrak{g} _{\ell } $ coincides (does not coincide) with the clockwise direction, the element that defines $ U^{\left( na \right) } _{f} $ is $ \mathfrak{g} _{\ell } $ ($ \left( \mathfrak{g} _{\ell } \right) ^{-1} $).}
			\end{figure}

			As a matter of fact, if we turn our attention only to the lattice gauge theories where there is no matter present (i.e., to the pure lattice gauge theories), it is not hard to demonstrate that this action can be written as the sum [\refcite{creutz,gauge-2}]
			\begin{equation}
				S = \sum _{f \in \mathcal{L} _{n}} \left[ \psi \left( U_{f} \right) + \psi \bigl( U^{-1} _{f} \bigr) \right] \label{action-S}
			\end{equation}
			of all the values of $ \psi \left( U_{f} \right) + \psi \bigl( U^{-1} _{f} \bigr) $ that can be estimated for the lattice faces, where $ U_{f} : G^{k} \rightarrow G $ is the holonomy associated with the $ f $-th lattice face, $ U^{-1} _{f} $ is its inverse\footnote{That is, if we consider, for instance, that $ U_{f} $ is calculated by using a counterclockwise orientation, $ U^{-1} _{f} $ must be interpreted as the holonomy that can be calculated by using a clockwise orientation.} and $ \psi : G \rightarrow \mathbb{C} $ is a class function. And as $ U_{f} $ is an application that, as Figure \ref{clockwise-holonomy} illustrates,
			can be defined as
			\begin{equation}
				U_{f} \left( \mathfrak{g} _{1} , \ldots , \mathfrak{g} _{k} \right) = \varphi \left( \ldots \varphi \left( \varphi \left( \mathfrak{g} _{1} , \mathfrak{g} _{2} \right) , \mathfrak{g} _{3} \right) , \ldots , \mathfrak{g} _{k} \right) = \mathfrak{g} _{f} \ , \label{holonomia}
			\end{equation}
			where [\refcite{birk}]
			\begin{equation}
				\left( \mathfrak{g} _{1} , \mathfrak{g} _{2} \right) \ \mapsto \ \varphi \left( \mathfrak{g} _{1} , \mathfrak{g} _{2} \right) = \mathfrak{g} \label{group-multiplication}
			\end{equation}
			is the application that gives a group structure to $ G $, it is not wrong to conclude that $ U_{f} $ is responsible for assigning an element of $ G $ with a lattice face $ f $ that has $ k $ sides. In other words, by assuming that
			\begin{itemize}
				\item there is a group element $ \mathfrak{g} _{\ell } $ associated with each lattice edge, and
				\item each lattice face can always be characterized by some sequence $ \mathfrak{g} _{1} , \ldots , \mathfrak{g} _{k} $ of these elements,
			\end{itemize}
			$ U_{f} $ allows us to multiply all elements of this sequence (in the same order they appear) and, therefore, assign a $ \mathfrak{g} _{f} $ to this $ f $-th lattice face\footnote{Note that, since $ U^{-1} _{f} $ must be interpreted as the inverse of $ U_{f} $, it is not hard to conclude that, if the latter is defined as $ U_{f} = \mathfrak{g} _{1} \cdot \mathfrak{g} _{2} \cdot \ldots \cdot \mathfrak{g} _{k-1} \cdot \mathfrak{g} _{k} $, the former should be defined as $ U^{-1} _{f} = \mathfrak{g} ^{-1} _{k} \cdot \mathfrak{g} ^{-1} _{k-1} \cdot \ldots \cdot \mathfrak{g} ^{-1} _{2} \cdot \mathfrak{g} ^{-1} _{1} $.}.
			
			\subsubsection{But why is action (\ref{action-S}) defined in this way?}
			
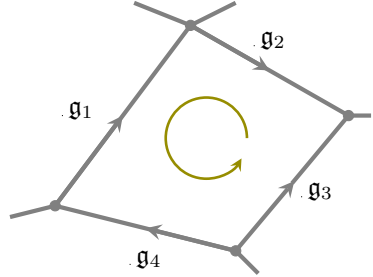
\begin{figure}[!t]
					\begin{center}
						\begin{tikzpicture}[
							scale=0.3,
							equation/.style={thin},
							trans/.style={color=gray, ultra thick},
							function/.style={->, color=gray, ultra thick, >=stealth}
							]
							\draw [color=olive,thin,->,line width=1.0pt,>=stealth] (24.5,3) arc[x radius=1.8cm, y radius =1.8cm, start angle=0, end angle=330];
							\draw[function,color=gray] (24,-2) -- (20,-1.0) node[above=2pt,right=-1pt] {};
							\draw[function,color=gray] (16,0) -- (19.0,4) node[above=2pt,right=-1pt] {};
							\draw[function,color=gray]  (24,-2) -- (26.6,1.1) node[above=2pt,right=-1pt] {};
							\draw[function,color=gray] (22,8) -- (25.5,6) node[above=2pt,right=-1pt] {};
							\draw[trans,color=gray] (16,0) -- (14,-0.5) node[above=2pt,right=-1pt] {};
							\draw[trans,color=gray] (24,-2) -- (25,-3) node[above=2pt,right=-1pt] {};
							\draw[trans,color=gray] (22,8) -- (24,9) node[above=2pt,right=-1pt] {};
							\draw[trans,color=gray] (22,8) -- (19.5,9) node[above=2pt,right=-1pt] {};
							\draw[trans,color=gray] (29,4) -- (30,4) node[above=2pt,right=-1pt] {};	
							\draw[trans,color=gray] (16,0) -- (24,-2) node[above=2pt,right=-1pt] {};
							\draw[trans,color=gray] (16,0) -- (22,8) node[above=2pt,right=-1pt] {};
							\draw[trans,color=gray] (22,8) -- (29,4) node[above=2pt,right=-1pt] {};
							\draw[trans,color=gray] (26.5,1.0) -- (29.0,4) node[above=2pt,right=-1pt] {};
							\draw[equation] (16.3,4.0) -- (16.3,4.0) node[above=2pt,right=-1pt] {$ \mathfrak{g} _{1} $};
							\draw[equation] (24.8,7.1) -- (24.8,7.1) node[above=2pt,right=-1pt] {$ \mathfrak{g} _{2} $};
							\draw[equation] (26.9,0.4) -- (26.9,0.4) node[above=2pt,right=-1pt] {$ \mathfrak{g} _{3} $};
							\draw[equation] (19.3,-2.6) -- (19.3,-2.6) node[above=2pt,right=-1pt] {$ \mathfrak{g} _{4} $};
							\draw[color=gray,fill=gray] (16,0) circle (1.5ex);
							\draw[color=gray,fill=gray] (24,-2) circle (1.5ex);
							\draw[color=gray,fill=gray] (22,8) circle (1.5ex);
							\draw[color=gray,fill=gray] (29,4) circle (1.5ex);
						\end{tikzpicture}
					\end{center}
					\caption{\protect\label{counterclockwise-holonomy} Here, we see the same lattice face as in Figure \ref{counterclockwise-holonomy}, but now it looks like we are calculating their holonomies by using the counterclockwise direction since the four possible results are \\ \hspace*{2.0cm} $ U^{\left( 1b \right) } _{f} = \left( \mathfrak{g} _{4} \right) ^{-1} \cdot \mathfrak{g} _{3} \cdot \left( \mathfrak{g} _{2} \right) ^{-1} \cdot \left( \mathfrak{g} _{1} \right) ^{-1} $ \ , \ $ U^{\left( 2b \right) } _{f} = \left( \mathfrak{g} _{1} \right) ^{-1} \cdot \left( \mathfrak{g} _{4} \right) ^{-1} \cdot \mathfrak{g} _{3} \cdot \left( \mathfrak{g} _{2} \right) ^{-1} $ \ , \\ \hspace*{2.0cm} $ U^{\left( 3b \right) } _{f} = \left( \mathfrak{g} _{2} \right) ^{-1} \cdot \left( \mathfrak{g} _{1} \right) ^{-1} \cdot \left( \mathfrak{g} _{4} \right) ^{-1} \cdot \mathfrak{g} _{3} $ \ \ and \ \ $ U^{\left( 4b \right) } _{f} = \mathfrak{g} _{3} \cdot \left( \mathfrak{g} _{2} \right) ^{-1} \cdot \left( \mathfrak{g} _{1} \right) ^{-1} \cdot \left( \mathfrak{g} _{4} \right) ^{-1} $ \ . \\ Note that, while it is clear that it is perfectly possible to calculate holonomies by taking clockwise and counterclockwise directions, it is important to observe that $ U^{\left( na \right) } _{f} = \bigl[ U^{\left( nb \right) } _{f} \bigr] ^{-1} $. That is, although it seems that we are getting eight different results for the holonomy around this lattice face, what we are getting are four different results and their inverses. Therefore, if we want to calculate the holonomies of all the faces of $ \mathcal{L} _{n} $, we must calculate the holonomy of a face by choosing a direction and apply this choice to all other faces in order to avoid confusing. Observe that this comment implies that the submanifold $ \mathcal{M} _{n} $ that $ \mathcal{L} _{n} $ (locally) discretizes must be orientable: after all, if $ \mathcal{M} _{n} $ is not orientable, this non-orientation will prevent us from choosing the direction in which these holonomies will be calculated.}
				\end{figure}

				In order to understand why (\ref{action-S}) is defined in this way, it is worth noting that, as illustrated in Figures \ref{clockwise-holonomy} and \ref{counterclockwise-holonomy},
				there is nothing to prevent $ U_{f} $ from being calculated in $ k $ different ways: in other words, there is nothing to prevent a face from being characterized by different elements of $ G $ when this gauge group is non-Abelian. In this fashion, as the gauge fields assigned to the lattice edges are responsible for performing parallel transports and, therefore, this concept of holonomy can be associated with an estimate of how curved is $ \mathcal{M} _{n} $ from the point of view of the faces of $ \mathcal{L} _{n} $ [\refcite{clarke}], there is a problem that needs to be fixed here. What problem? The action (\ref{action-S}) needs to be defined so that the Boltzmann factor of each lattice face is invariant. Thus, by observing that every class function $ \psi $ is such that
				\begin{equation*}
					\psi \left( \mathfrak{g} ^{\prime } \cdot \mathfrak{g} ^{\prime \prime } \cdot \mathfrak{g} ^{\prime \prime \prime } \right) = \psi \left( \mathfrak{g} ^{\prime \prime } \cdot \mathfrak{g} ^{\prime \prime \prime } \cdot \mathfrak{g} ^{\prime } \right) = \psi \left( \mathfrak{g} ^{\prime \prime \prime } \cdot \mathfrak{g} ^{\prime } \cdot \mathfrak{g} ^{\prime \prime } \right)
				\end{equation*}
				holds for any group elements $ \mathfrak{g} ^{\prime } $, $ \mathfrak{g} ^{\prime \prime } $ and $ \mathfrak{g} ^{\prime \prime \prime } $, it is precisely the use of a class function that leads to one action (\ref{action-S}) that does not depend on these $ k $ possible choices to calculate $ U_{f} $. Note that, since $ \psi \left( U_{f} \right) $ is a complex number, the superposition $ \psi \left( U_{f} \right) + \psi \bigl( U^{-1} _{f} \bigr) $ defines a real number.
				
			\subsubsection{A small parenthesis}
				
				Of course, what we have just said still does not clarify all aspects of why (\ref{action-S}) is defined in this way. But, before we delve a little further into this discussion, it is important to make a little parenthesis here to make three important observations. And the first one is related to the fact that this description of lattice gauge theories, based on the use of this partition function (\ref{pre-part-function}), had its origin in Ref. [\refcite{wegner}], where a generalization of the Ising model [\refcite{ising}] was presented by assigning the spin variables to the lattice edges. After all, as the use of this generalization was successful in a first work, where it was possible to evaluate questions related to the quark confinement [\refcite{wilson-loops}], the prototype of this generalization was used as a cornerstone for the development of other lattice gauge theories.
				
				By putting that historical information aside, which justifies the successful use of (\ref{pre-part-function}) in the formulation of lattice gauge theories is that, for example, it provides a good way to compute the path integral over all the possible configurations of the gauge fields in a quantum field theory [\refcite{mackenzie}]. In other words, just as the path integral is a sum over all possible histories of a system [\refcite{feynman}], this partition function (\ref{pre-part-function}) computes this sum in a more tractable way by breaking it down into a product of simpler factors. We will return to this discussion later on, when we are finished clarifying why (\ref{action-S}) is defined in this way. For now, the second observation we need to make here concerns the fact that, in the case of the lattice gauge theories where there is matter present in the vertices, their actions are given by [\refcite{fradkin}]
				\begin{equation}
					\beta S = \beta ^{\prime } S_{\mathrm{gauge}} + \beta ^{\prime \prime } S_{\mathrm{matter}} \ , \label{more-general-action-S}
				\end{equation}
				where $ \beta ^{\prime } $ and $ \beta ^{\prime \prime } $ are two real numbers, $ S_{\mathrm{gauge}} $ is the same action (\ref{action-S}), and the new $ S_{\mathrm{matter}} $ must describe how the matter fields interact with each other. But when we note that this historical information make it clear, for example, that these more general lattice gauge theories need to bring the Ising models as special cases, the interaction model between first neighbours endorses that [\refcite{fradkin,seiler}]
				\begin{equation}
					S_{\mathrm{matter}} = \sum _{\ell \in \mathcal{L} _{n}} \left\langle v^{\left( \ell \right) } _{1} , \rho \left( \mathfrak{g} _{\ell } \right) \cdot v^{\left( \ell \right) } _{2} \right\rangle \ , \label{matter-action}
				\end{equation}
				since the term inside this summation is an inner product that models how aligned are the two matter fields $ v^{\left( \ell \right) } _{1} $ and $ v^{\left( \ell \right) } _{2} $ that endpoint the $ \ell $-th lattice edge\footnote{Note that this sum is similar to the one in (\ref{action-S}): i.e., as the symbol \textquotedblleft $ \ell $\textquotedblright \hspace*{0.01cm} indexes the $ \ell $-th edge of $ \mathcal{L} _{n} $, this action $ S_{\mathrm{matter}} $ sums all the values that $ \left\langle v^{\left( \ell \right) } _{1} , \rho \left( \mathfrak{g} _{\ell } \right) \cdot v^{\left( \ell \right) } _{2} \right\rangle $ assumes for all the lattice edges.}. It is clear that the presence of $ \rho \left( \mathfrak{g} _{\ell } \right) $ makes this inner product in (\ref{matter-action}) a little different from those that define the Ising and Potts models [\refcite{potts,wu}]. And what explains this difference is the fact that, as these matter fields $ v^{\left( \ell \right) } _{1} $ and $ v^{\left( \ell \right) } _{2} $ need to interact with each other, this interaction needs to be moderated by the lattice gauge field that appear on the $ \ell $-edge. Thus, by remembering that a group always admits a matrix representation $ \rho $, it is correct to say that the product $ \rho \left( \mathfrak{g} _{\ell } \right) \cdot v^{\left( \ell \right) } _{2} $, between the matrix $ \rho \left( \mathfrak{g} _{\ell } \right) $ and the vector $ v^{\left( \ell \right) } _{2} $ (which belongs to a finite-dimensional vector space), can be interpreted as a gauge group action [\refcite{james}]. That is, these matter fields are coupled to the lattice gauge fields by using this group action in a situation where $ \beta ^{\prime \prime } $ is a non-zero real number [\refcite{fradkin}]. % aligned from the µ point of view.
				
				However, while this second observation suggests that the lattice gauge theories that we need to evaluate are those whose actions are more general than (\ref{action-S}), we still need to make the third (and, perhaps, most important) observation here. After all, regardless of the form that $ S_{\mathrm{matter}} $ takes, (\ref{pre-part-function}) and (\ref{more-general-action-S}) allow us to conclude that, if a pure lattice gauge theory can be interpreted as a Hamiltonian system with constraints, the pure gauge action (\ref{action-S}) alone can already lead to the formulation where these first-class constraints appear. In other words, as
				\begin{itemize}
					\item the action (\ref{more-general-action-S}) is such that
					\begin{equation*}
						e^{- \left( \beta ^{\prime } S_{\mathrm{gauge}} + \beta ^{\prime \prime } S_{\mathrm{matter}} \right) } = e^{- \beta ^{\prime } S_{\mathrm{gauge}}} \cdot e^{- \beta ^{\prime \prime } S_{\mathrm{matter}}} , \quad \textnormal{and}
					\end{equation*}
					\item the Hamiltonian formulation is obtained by taking the logarithm of (\ref{pre-part-function}),
				\end{itemize}
				if a pure lattice gauge theory can be interpreted as a Hamiltonian system with constraints, a more general lattice gauge theory can also be interpreted in the same way. Note that an alternative way to understand this same conclusion is, for example, by exploring the limiting case where $ v^{\left( \ell \right) } _{1} $ and $ v^{\left( \ell \right) } _{2} $ belong to a one-dimensional vector space: as, in this limiting case, $ \rho \left( \mathfrak{g} _{\ell } \right) \cdot v^{\left( \ell \right) } _{1,2} = v^{\left( \ell \right) } _{1,2} $ and, therefore,
				\begin{equation}
					\left\langle v^{\left( \ell \right) } _{1} , \rho \left( \mathfrak{g} _{\ell } \right) \cdot v^{\left( \ell \right) } _{2} \right\rangle = 1
				\end{equation}
				holds for all the values of $ \ell $, the advent of the correspondence principle in Physics ensures that this interpretation, of a more general lattice gauge theory as a Hamiltonian system with constraints, is the sole responsibility of $ S_{\mathrm{gauge}} $, since the presence of these matter fields on the lattice vertices is irrelevant for this purpose.
			
		\subsubsection{\label{gauge-subsection} Gauge transformations}
		
			Having made this small parenthesis, now we can return to the discussion of why (\ref{action-S}) is defined in this way, and we will do this by explaining why these lattice systems, where there is no matter present, are usually interpreted as gauge theories. And by remembering that the characterization of any physical system as a gauge theory is directly related to the \emph{covariance} of its equations of motion [\refcite{ait-hey-1}], there are two critical remarks that we should make here, and the first one is precisely related to this concept of covariance. After all, although it is quite common to \textquotedblleft hear\textquotedblright \hspace*{0.01cm} that the equations of motion of a physical system are covariant because they maintain their \textquotedblleft form\textquotedblright \hspace*{0.01cm} unchanged under gauge transformations, the truth is that the origin of this predicate \textquotedblleft covariance\textquotedblright \hspace*{0.01cm} is geometric: i.e., this predicate reflects the fact that these equations depend exclusively on the parameters/functions that intrinsically describe the geometry of this physical system\footnote{In other words, the equations of motion
			\begin{equation*}
				\dot{z} = \left\{ z , H_{T} \left( z \right) \right\} _{\Phi \left( z \right) = 0}
			\end{equation*}
			of a physical system are rewritten as
			\begin{equation*}
				\dot{z} ^{\prime } = \left\{ z^{\prime } , H^{\prime } _{T} \left( z^{\prime } \right) \right\} _{\Phi \left( z^{\prime } \right) = 0}
			\end{equation*}
			under a gauge transformation $ z \mapsto z^{\prime } = \mathcal{T} \left( z \right) $ [\refcite{gitman}]. Note that this reinforces the comment we made in the Introduction, about the fact that the physics of a classical gauge system can be described by (\ref{eq-physics}). After all, as $ \omega $ describes the intrinsic parameters of $ T^{\ast } \mathcal{M} _{n} $, the gauge transformations $ \left( \omega , \Omega \right) \mapsto \left( \omega , \Omega ^{\prime } \right) = \mathcal{T} \left( \omega , \Omega \right) $ never change (\ref{eq-physics}) [\refcite{gitman,man-gd}].} [\refcite{man-gr,man-gd}]. And why is this first remark critical? Because, as these lattice gauge theories are usually described without ever mentioning that $ \mathcal{L} _{n} $ discretizes a submanifold, it is not wrong to assert that, at least, the elements of the gauge group $ G $ allow to identify the intrinsic parameters of $ \mathcal{L} _{n} $.
			
			But what does it mean to say that the equations of motion of a lattice gauge theory are covariant for someone looking, for instance, only at the configuration of the gauge elements assigned to the lattice edges? As we have already said that (\ref{action-S}) needs to be defined so that the Boltzmann factor of each lattice face is invariant, the natural answer to this question is: the equations of motion of a lattice gauge theory are covariant under lattice gauge transformations that do not change the holonomies around the lattice faces, and this is precisely the second critical remark that we needed to make. And as we have already said that $ U_{f} $ can be associated with an estimate of how curved is $ \mathcal{M} _{n} $ from the point of view of the faces of $ \mathcal{L} _{n} $, it is not difficult to conclude that: saying that the lattice gauge transformations are those that do not change these holonomies is equivalent to saying that, whatever the new group elements that will be assigned to the lattice edges, these new group elements continue to intrinsically describe the same $ \mathcal{L} _{n} $. Therefore, by remembering that the elements of a non-trivial group cannot be unequivocally expressed as a product of $ k $ group elements [\refcite{roy}], the characterization of these lattice systems as gauge theories can be related to the freedom we have to change any of the group elements $ \mathfrak{g} _{\ell } $, which are assigned to the edges of $ \mathcal{L} _{n} $, to other $ \mathfrak{g} ^{\prime } _{\ell } $ as long as the value of $ U_{f} $ remains unchanged.
			
			Given these two critical remarks, we can conclude that, if we know anyone of the field configurations of a pure gauge lattice system, all the others can be obtained from this first one through lattice gauge transformations. Yet, as any lattice edge may belong to more than one face of $ \mathcal{L} _{n} $, a good way to perform these transformations is by modifying all the group elements, which are assigned to all the $ n_{v} $ edges that composes a same lattice vertex $ v $, for others [\refcite{oeckl,bahr}] \label{gauge-transformation}
			\begin{itemize}
				\item[(a)] $ \mathfrak{g} \cdot \mathfrak{g} _{\ell } $, if the $ \ell $-th edge orientation pointing out of this $ v $-th vertex, or
				\item[(b)] $ \mathfrak{g} _{\ell } \cdot \mathfrak{g} ^{-1} $, otherwise.
			\end{itemize}
			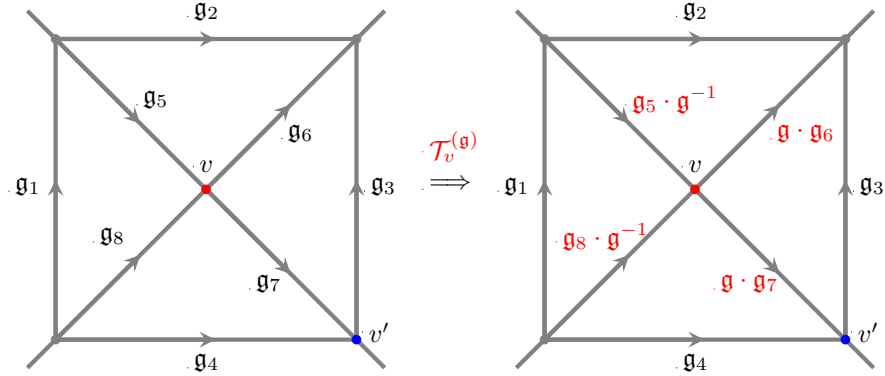
\begin{figure}[!t]
				\begin{center}
					\begin{tikzpicture}[
						scale=0.25,
						equation/.style={thin},
						trans/.style={color=gray, ultra thick},
						function/.style={->, color=gray, ultra thick, >=stealth}
						]
						\draw[function,color=gray] (0,0) -- (8.5,0) node[above=2pt,right=-1pt] {};
						\draw[function,color=gray] (0,0) -- (0,8.5) node[above=2pt,right=-1pt] {};
						\draw[function,color=gray] (16.0,0) -- (16.0,8.5) node[above=2pt,right=-1pt] {};
						\draw[function,color=gray] (0,16.0) -- (8.5,16.0) node[above=2pt,right=-1pt] {};
						\draw[function,color=gray] (0,0) -- (4.5,4.5) node[above=2pt,right=-1pt] {};
						\draw[function,color=gray] (8.0,8.0) -- (12.5,12.5) node[above=2pt,right=-1pt] {};
						\draw[function,color=gray] (0,16.0) -- (4.5,11.5) node[above=2pt,right=-1pt] {};
						\draw[function,color=gray] (8.0,8.0) -- (12.5,3.5) node[above=2pt,right=-1pt] {};
						\draw[trans,color=gray] (0,0) -- (16.0,0) node[above=2pt,right=-1pt] {};
						\draw[trans,color=gray] (16.0,0) -- (16.0,16.0) node[above=2pt,right=-1pt] {};
						\draw[trans,color=gray] (16.0,16.0) -- (0,16.0) node[above=2pt,right=-1pt] {};
						\draw[trans,color=gray] (0,0) -- (0,16.0) node[above=2pt,right=-1pt] {};
						\draw[trans,color=gray] (0,0) -- (8.0,8.0) node[above=2pt,right=-1pt] {};
						\draw[trans,color=gray] (16.0,0) -- (8.0,8.0) node[above=2pt,right=-1pt] {};
						\draw[trans,color=gray] (16.0,16.0) -- (8.0,8.0) node[above=2pt,right=-1pt] {};
						\draw[trans,color=gray] (0,16.0) -- (8.0,8.0) node[above=2pt,right=-1pt] {};
						\draw[trans,color=gray] (0,0) -- (-1.5,-1.5) node[above=2pt,right=-1pt] {};
						\draw[trans,color=gray] (16.0,0) -- (17.5,-1.5) node[above=2pt,right=-1pt] {};
						\draw[trans,color=gray] (16.0,16.0) -- (17.5,17.5) node[above=2pt,right=-1pt] {};
						\draw[trans,color=gray] (0,16.0) -- (-1.5,17.5) node[above=2pt,right=-1pt] {};
						\draw[equation] (-2.5,7.9) -- (-2.5,7.9) node[above=2pt,right=-1pt] {$ \mathfrak{g} _{1} $};
						\draw[equation] (7.0,17.2) -- (7.0,17.2) node[above=2pt,right=-1pt] {$ \mathfrak{g} _{2} $};
						\draw[equation] (16.4,7.9) -- (16.4,7.9) node[above=2pt,right=-1pt] {$ \mathfrak{g} _{3} $};
						\draw[equation] (7.0,-1.5) -- (7.0,-1.5) node[above=2pt,right=-1pt] {$ \mathfrak{g} _{4} $};
						\draw[equation] (4.3,12.5) -- (4.3,12.5) node[above=2pt,right=-1pt] {$ \mathfrak{g} _{5} $};
						\draw[equation] (12.0,10.7) -- (12.0,10.7) node[above=2pt,right=-1pt] {$ \mathfrak{g} _{6} $};
						\draw[equation] (10.3,2.7) -- (10.3,2.7) node[above=2pt,right=-1pt] {$ \mathfrak{g} _{7} $};
						\draw[equation] (2.0,5.2) -- (2.0,5.2) node[above=2pt,right=-1pt] {$ \mathfrak{g} _{8} $};
						\draw[color=gray,fill=gray] (0,0) circle (1.5ex);
						\draw[color=blue,fill=blue] (16,0) circle (1.5ex);
						\draw[color=gray,fill=gray] (0,16) circle (1.5ex);
						\draw[color=gray,fill=gray] (16,16) circle (1.5ex);
						\draw[color=red,fill=red] (8,8) circle (1.5ex);
						\draw[equation] (7.3,9.2) -- (7.3,9.2) node[above=0pt,right=-1pt] {$ v $};
						\draw[equation] (16.3,0.4) -- (16.3,0.4) node[above=0pt,right=-1pt] {$ v^{\prime } $};
						\draw[equation,color=red] (19.5,10.0) -- (19.5,10.0) node[above=2pt,right=-1pt] {$ \mathcal{T} ^{\left( \mathfrak{g} \right) } _{v} $};
						\draw[equation] (19.5,8.0) -- (19.5,8.0) node[above=2pt,right=-1pt] {$ \Longrightarrow $};
						\draw[function,color=gray] (26.0,0) -- (34.5,0) node[above=2pt,right=-1pt] {};
						\draw[function,color=gray] (26.0,0) -- (26.0,8.5) node[above=2pt,right=-1pt] {};
						\draw[function,color=gray] (42.0,0) -- (42.0,8.5) node[above=2pt,right=-1pt] {};
						\draw[function,color=gray] (26.0,16.0) -- (34.5,16.0) node[above=2pt,right=-1pt] {};
						\draw[function,color=gray] (26.0,0) -- (30.5,4.5) node[above=2pt,right=-1pt] {};
						\draw[function,color=gray] (34.0,8.0) -- (38.5,12.5) node[above=2pt,right=-1pt] {};
						\draw[function,color=gray] (26.0,16.0) -- (30.5,11.5) node[above=2pt,right=-1pt] {};
						\draw[function,color=gray] (34.0,8.0) -- (38.5,3.5) node[above=2pt,right=-1pt] {};
						\draw[trans,color=gray] (26.0,0) -- (42.0,0) node[above=2pt,right=-1pt] {};
						\draw[trans,color=gray] (42.0,0) -- (42.0,16.0) node[above=2pt,right=-1pt] {};
						\draw[trans,color=gray] (42.0,16.0) -- (26.0,16.0) node[above=2pt,right=-1pt] {};
						\draw[trans,color=gray] (26.0,0) -- (26.0,16.0) node[above=2pt,right=-1pt] {};
						\draw[trans,color=gray] (26.0,0) -- (34.0,8.0) node[above=2pt,right=-1pt] {};
						\draw[trans,color=gray] (42.0,0) -- (34.0,8.0) node[above=2pt,right=-1pt] {};
						\draw[trans,color=gray] (42.0,16.0) -- (34.0,8.0) node[above=2pt,right=-1pt] {};
						\draw[trans,color=gray] (26.0,16.0) -- (34.0,8.0) node[above=2pt,right=-1pt] {};
						\draw[trans,color=gray] (26.0,0) -- (24.5,-1.5) node[above=2pt,right=-1pt] {};
						\draw[trans,color=gray] (42.0,0) -- (43.5,-1.5) node[above=2pt,right=-1pt] {};
						\draw[trans,color=gray] (42.0,16.0) -- (43.5,17.5) node[above=2pt,right=-1pt] {};
						\draw[trans,color=gray] (26.0,16.0) -- (24.5,17.5) node[above=2pt,right=-1pt] {};
						\draw[equation] (23.5,7.9) -- (23.5,7.9) node[above=2pt,right=-1pt] {$ \mathfrak{g} _{1} $};
						\draw[equation] (33.0,17.2) -- (33.0,17.2) node[above=2pt,right=-1pt] {$ \mathfrak{g} _{2} $};
						\draw[equation] (42.4,7.9) -- (42.4,7.9) node[above=2pt,right=-1pt] {$ \mathfrak{g} _{3} $};
						\draw[equation] (33.0,-1.5) -- (33.0,-1.5) node[above=2pt,right=-1pt] {$ \mathfrak{g} _{4} $};
						\draw[equation,color=red] (30.3,12.5) -- (30.3,12.5) node[above=2pt,right=-1pt] {$ \mathfrak{g} _{5} \cdot \mathfrak{g} ^{-1} $};
						\draw[equation,color=red] (38.0,10.7) -- (38.0,10.7) node[above=2pt,right=-1pt] {$ \mathfrak{g} \cdot \mathfrak{g} _{6} $};
						\draw[equation,color=red] (35.0,2.7) -- (35.0,2.7) node[above=2pt,right=-1pt] {$ \mathfrak{g} \cdot \mathfrak{g} _{7} $};
						\draw[equation,color=red] (26.6,5.2) -- (26.6,5.2) node[above=2pt,right=-1pt] {$ \mathfrak{g} _{8} \cdot \mathfrak{g} ^{-1} $};
						\draw[color=gray,fill=gray] (26,0) circle (1.5ex);
						\draw[color=blue,fill=blue] (42,0) circle (1.5ex);
						\draw[color=gray,fill=gray] (26,16) circle (1.5ex);
						\draw[color=gray,fill=gray] (42,16) circle (1.5ex);
						\draw[color=red,fill=red] (34,8) circle (1.5ex);
						\draw[equation] (33.3,9.2) -- (33.3,9.2) node[above=0pt,right=-1pt] {$ v $};
						\draw[equation] (42.3,0.4) -- (42.3,0.4) node[above=0pt,right=-1pt] {$ v^{\prime } $};
					\end{tikzpicture}
				\end{center}
				\caption{\protect\label{gauge-transformation-example} On the left, we can see a piece of a $ 2 $-dimensional lattice $ \mathcal{L} _{2} $ whose edges support elements of the gauge group $ G $. Note that, in this piece, we have four faces/edges sharing the same vertex $ v $ (highlighted in red colour). On the right, we see the result of a lattice gauge transformation $ \mathcal{T} ^{\left( \mathfrak{g} \right) } _{v} $ (highlighted with the same red colour), which was performed on the group elements that are assigned to these four edges. And as the discussion presented in Figures \ref{clockwise-holonomy} and \ref{counterclockwise-holonomy} claims that we must choose a single direction to calculate the face holonomies of this piece, it is not difficult to conclude that $ \mathcal{T} ^{\left( \mathfrak{g} \right) } _{v} $ does not modify these holonomies.}
			\end{figure}	
			And an instructive example of these lattice gauge transformations $ \mathcal{T} ^{\left( \mathfrak{g} \right) } _{v} : G^{n_{v}} \rightarrow G ^{n_{v}} $ can be seen in Figure \ref{gauge-transformation-example}: after all, if we calculate all the holonomies by using the same counterclockwise orientation, it is not difficult to see that $ \mathcal{T} ^{\left( \mathfrak{g} \right) } _{v} $ does not change the holonomies around the lattice faces. Observe that, as
			\begin{equation*}
				\psi \left( U_{f} \right) = \psi \left( \mathfrak{g} ^{-1} \cdot \mathfrak{g} \cdot U_{f} \right) = \psi \left( \mathfrak{g} \cdot U_{f} \cdot \mathfrak{g} ^{-1} \right)
			\end{equation*}
			holds for all $ \mathfrak{g} \in G $, this no-change in the values of $ \psi \left( U_{f} \right) $ can be associated with the fact that $ U_{f} $ and $ \mathfrak{g} \cdot U_{f} \cdot \mathfrak{g} ^{-1} $ belong to the same conjugacy class [\refcite{james}].
			
\begin{figure}[!t]
				\begin{center}
					\begin{tikzpicture}[
						scale=0.24,
						equation/.style={thin},
						trans/.style={color=gray, ultra thick},
						function/.style={->, color=gray, ultra thick, >=stealth}
						]
						\draw[function,color=gray] (0,0) -- (8.5,0) node[above=2pt,right=-1pt] {};
						\draw[function,color=gray] (16.0,0) -- (16.0,8.5) node[above=2pt,right=-1pt] {};
						\draw[function,color=gray] (0,0) -- (4.5,4.5) node[above=2pt,right=-1pt] {};
						\draw[function,color=gray] (8.0,8.0) -- (12.5,12.5) node[above=2pt,right=-1pt] {};
						\draw[function,color=gray] (8.0,8.0) -- (12.5,3.5) node[above=2pt,right=-1pt] {};
						\draw[trans,color=gray] (0,0) -- (16.0,0) node[above=2pt,right=-1pt] {};
						\draw[trans,color=gray] (16.0,0) -- (16.0,16.0) node[above=2pt,right=-1pt] {};
						\draw[trans,color=gray] (0,0) -- (8.0,8.0) node[above=2pt,right=-1pt] {};
						\draw[trans,color=gray] (16.0,0) -- (8.0,8.0) node[above=2pt,right=-1pt] {};
						\draw[trans,color=gray] (16.0,16.0) -- (8.0,8.0) node[above=2pt,right=-1pt] {};
						\draw[trans,color=gray] (0,0) -- (-1.5,-1.5) node[above=2pt,right=-1pt] {};
						\draw[trans,color=gray] (16.0,0) -- (17.5,-1.5) node[above=2pt,right=-1pt] {};
						\draw[trans,color=gray] (16.0,16.0) -- (17.5,17.5) node[above=2pt,right=-1pt] {};
						\draw[trans,color=gray] (6.5,9.5) -- (8.0,8.0) node[above=2pt,right=-1pt] {};
						\draw[function,color=gray] (16,0) -- (20.5,-4.5) node[above=2pt,right=-1pt] {};
						\draw[trans,color=gray] (16.0,0) -- (24.0,-8.0) node[above=2pt,right=-1pt] {};
						\draw[equation] (16.4,7.9) -- (16.4,7.9) node[above=2pt,right=-1pt] {$ \mathfrak{g} _{3} $};
						\draw[equation] (7.0,-1.5) -- (7.0,-1.5) node[above=2pt,right=-1pt] {$ \mathfrak{g} _{4} $};
						\draw[equation,color=red] (12.0,10.7) -- (12.0,10.7) node[above=2pt,right=-1pt] {$ \mathfrak{g} \cdot \mathfrak{g} _{6} $};
						\draw[equation,color=red] (9.0,2.7) -- (9.0,2.7) node[above=2pt,right=-1pt] {$ \mathfrak{g} \cdot \mathfrak{g} _{7} $};
						\draw[equation,color=red] (0.6,5.2) -- (0.6,5.2) node[above=2pt,right=-1pt] {$ \mathfrak{g} _{8} \cdot \mathfrak{g} ^{-1} $};
						\draw[equation] (18.0,-5.0) -- (18.0,-5.0) node[above=2pt,right=-1pt] { $ \mathfrak{g} _{9} $};
						\draw[color=gray,fill=gray] (0,0) circle (1.5ex);
						\draw[color=blue,fill=blue] (16,0) circle (1.5ex);
						\draw[color=gray,fill=gray] (16,16) circle (1.5ex);
						\draw[color=red,fill=red] (8,8) circle (1.5ex);
						\draw[equation] (7.3,9.2) -- (7.3,9.2) node[above=0pt,right=-1pt] {$ v $};
						\draw[equation] (16.3,0.4) -- (16.3,0.4) node[above=0pt,right=-1pt] {$ v^{\prime } $};
						\draw[equation,color=blue] (21.0,10.0) -- (21.0,10.0) node[above=2pt,right=-1pt] {$ \mathcal{T} ^{\left( \mathfrak{g} ^{\prime } \right) } _{v^{\prime }} $};
						\draw[equation] (21.0,8.0) -- (21.0,8.0) node[above=2pt,right=-1pt] {$ \Longrightarrow $};
						\draw[function,color=gray] (25.0,0) -- (33.5,0) node[above=2pt,right=-1pt] {};
						\draw[function,color=gray] (41.0,0) -- (41.0,8.5) node[above=2pt,right=-1pt] {};
						\draw[function,color=gray] (25.0,0) -- (29.5,4.5) node[above=2pt,right=-1pt] {};
						\draw[function,color=gray] (33.0,8.0) -- (37.5,12.5) node[above=2pt,right=-1pt] {};
						\draw[function,color=gray] (33.0,8.0) -- (37.5,3.5) node[above=2pt,right=-1pt] {};
						\draw[trans,color=gray] (25.0,0) -- (41.0,0) node[above=2pt,right=-1pt] {};
						\draw[trans,color=gray] (41.0,0) -- (41.0,16.0) node[above=2pt,right=-1pt] {};
						\draw[trans,color=gray] (25.0,0) -- (33.0,8.0) node[above=2pt,right=-1pt] {};
						\draw[trans,color=gray] (41.0,0) -- (33.0,8.0) node[above=2pt,right=-1pt] {};
						\draw[trans,color=gray] (41.0,16.0) -- (33.0,8.0) node[above=2pt,right=-1pt] {};
						\draw[trans,color=gray] (25.0,0) -- (23.5,-1.5) node[above=2pt,right=-1pt] {};
						\draw[trans,color=gray] (41.0,0) -- (42.5,-1.5) node[above=2pt,right=-1pt] {};
						\draw[trans,color=gray] (41.0,16.0) -- (42.5,17.5) node[above=2pt,right=-1pt] {};
						\draw[trans,color=gray] (31.5,9.5) -- (33,8) node[above=2pt,right=-1pt] {};
						\draw[function,color=gray] (41,0) -- (45.5,-4.5) node[above=2pt,right=-1pt] {};
						\draw[trans,color=gray] (41,0) -- (49.0,-8.0) node[above=2pt,right=-1pt] {};
						\draw[equation,color=blue] (41.4,7.9) -- (41.4,7.9) node[above=2pt,right=-1pt] {$ \mathfrak{g} ^{\prime } \cdot \mathfrak{g} _{3} $};
						\draw[equation,color=blue] (31.0,-1.5) -- (31.0,-1.5) node[above=2pt,right=-1pt] {$ \mathfrak{g} _{4} \cdot \left( \mathfrak{g} ^{\prime } \right) ^{-1} $};
						\draw[equation,color=red] (37.0,10.7) -- (37.0,10.7) node[above=2pt,right=-1pt] {$ \mathfrak{g} \cdot \mathfrak{g} _{6} $};
						\draw[equation,color=blue] (29.6,2.3) -- (29.6,2.3) node[above=2pt,right=-1pt] {$ \mathfrak{g} \cdot \mathfrak{g} _{7} \cdot \left( \mathfrak{g} ^{\prime } \right) ^{-1} $};
						\draw[equation,color=red] (25.6,5.2) -- (25.6,5.2) node[above=2pt,right=-1pt] {$ \mathfrak{g} _{8} \cdot \mathfrak{g} ^{-1} $};
						\draw[equation,color=blue] (41.3,-5.0) -- (41.3,-5.0) node[above=2pt,right=-1pt] { $ \mathfrak{g} ^{\prime } \cdot \mathfrak{g} _{9} $};
						\draw[color=gray,fill=gray] (25,0) circle (1.5ex);
						\draw[color=blue,fill=blue] (41,0) circle (1.5ex);
						\draw[color=gray,fill=gray] (41,16) circle (1.5ex);
						\draw[color=red,fill=red] (33,8) circle (1.5ex);
						\draw[equation] (32.3,9.2) -- (32.3,9.2) node[above=0pt,right=-1pt] {$ v $};
						\draw[equation] (41.3,0.4) -- (41.3,0.4) node[above=0pt,right=-1pt] {$ v^{\prime } $};
					\end{tikzpicture}
				\end{center}
				\caption{\protect\label{transporter-example} Although Figure \ref{gauge-transformation-example} shows us a single example of a lattice gauge transformation, it is worth noting that not only this example, but also the way that we introduce the concept of the lattice gauge transformation are equivalent to the definition given in the literature. And in order to understand the reason of this equivalence, it is enough to observe that, when we perform a new transformation $ \mathcal{T} ^{\left( \mathfrak{g} ^{\prime } \right) } _{v^{\prime }} $ on the same piece shown in Figure \ref{gauge-transformation-example}, but now on the group elements assigned to the edges that share the $ v^{\prime } $-th vertex, the element $ \mathfrak{g} \cdot \mathfrak{g} _{7} $ gives way to $ \mathfrak{g} \cdot \mathfrak{g} _{7} \cdot \left( \mathfrak{g} ^{\prime } \right) ^{-1} $. After all, by remembering that $ \mathcal{T} ^{\left( \mathfrak{g} ^{\prime } \right) } _{v^{\prime }} \circ \mathcal{T} ^{\left( \mathfrak{g} \right) } _{v} $ is, for instance, a lattice gauge transformation when $ \mathcal{T} ^{\left( \mathfrak{g} \right) } _{v} $ and $ \mathcal{T} ^{\left( \mathfrak{g} ^{\prime } \right) } _{v^{\prime }} $ are also, the fact that $ \mathfrak{g} _{7} \mapsto \mathfrak{g} \cdot \mathfrak{g} _{7} \cdot \left( \mathfrak{g} ^{\prime } \right) ^{-1} $ retrieves the definition (of lattice gauge transformations) given in the literature [\protect\refcite{oeckl}] makes it clear that $ \mathcal{T} ^{\left( \mathfrak{g} \right) } _{v} $, $ \mathcal{T} ^{\left( \mathfrak{g} ^{\prime } \right) } _{v^{\prime }} $ and, therefore, $ \mathcal{T} ^{\left( \mathfrak{g} ^{\prime } \right) } _{v^{\prime }} \circ \mathcal{T} ^{\left( \mathfrak{g} \right) } _{v} $ can actually be interpreted as such.}
			\end{figure}
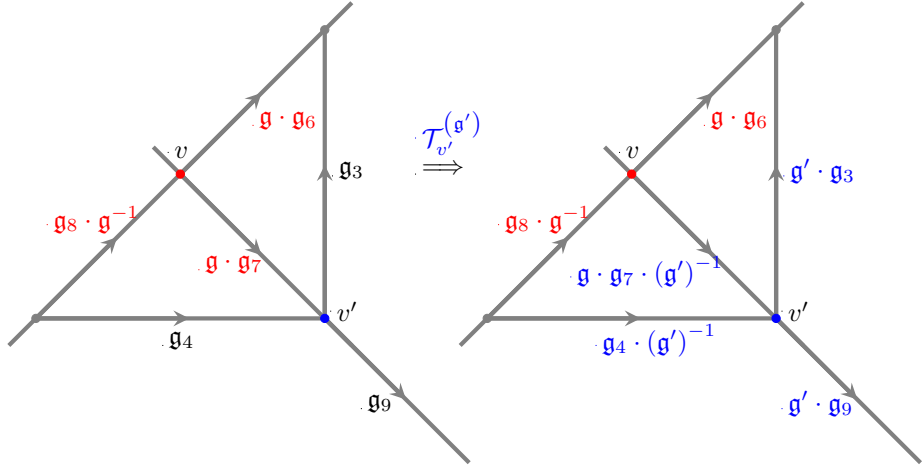

			Anyway, although Figure \ref{transporter-example}
			continues to explore the same situation as Figure \ref{gauge-transformation-example} to, for example, reinforce that the way that we introduce the concept of the lattice gauge transformation here is equivalent to the definition given in the literature, it is undeniable that everything we have said so far has made one thing very clear: the class function $ \psi $ is the main protagonist of the action (\ref{action-S}). After all, as this function
			\begin{itemize}
				\item gets its name because it defines different conjugacy classes by assigning a distinct complex number to each one of them, and (consequently)
				\item gathers all the possible holonomies, which can be calculated for the same lattice face, in the same conjugacy class,
			\end{itemize}
			it is precisely its presence in (\ref{action-S}) that makes this action invariant under these lattice gauge transformations. In this fashion, since these lattice gauge transformations cannot change the physics of this lattice system, it is correct to say that the real explanation, for the fact that (\ref{pre-part-function}) can be written by using (\ref{action-S}), is directly associated with the fact that the presence of $ \psi $ in (\ref{action-S}) allows us to define (\ref{pre-part-function}) as a sum over all the gauge-invariant configurations. In plain English, in the same way that the partition function (\ref{pre-part-function}) plays a role analogous to that of the path integral, the use of this action (\ref{action-S}) comes to establish a covariance in these lattice gauge theories because, as we just said, their gauge transformations cannot change the physics of this lattice system. %%%%%% 89B 
			
	\section{\label{constrained-characterisation} The Hamiltonian formulation with constraints}
		
		Although it has become clear that the interpretation of a pure lattice gauge theory is closely associated with the fact that, for example, $ \mathcal{T} ^{\left( \mathfrak{g} \right) } _{v} $ does not modify the holonomies around the lattice faces, it is interesting to note that (\ref{action-S}) and, consequently, (\ref{pre-part-function}) make no mention of the fact that $ \mathcal{L} _{n} $ can be interpreted as the lattice that discretizes a submanifold. So, since the elements of $ G $ lead us to an intrinsic parameterization of $ \mathcal{L} _{n} $ through of $ \psi $ (because to gather all the possible face holonomies in the same conjugacy class reduces the number of degrees of freedom in this lattice system), the burning question that we need to answer now is: how can we find a partition function that not only makes it clear that $ \mathcal{L} _{n} $ can be interpreted as the lattice that discretizes a submanifold, but also allows us to recover (\ref{pre-part-function}) by taking $ \Phi _{f} = 0 $?
		
		By analogy with what was said in the Introduction, a naive answer that we can give to this burning question is: by finding a way to find what are the extrinsic parameters to $ \mathcal{L} _{n} $. And in order to find these extrinsic parameters, it seems to be interesting to assume that $ G $ is a subgroup of $ G^{\prime } $ because, as the elements of $ G $ lead us to an intrinsic parameterization of $ \mathcal{L} _{n} $, the elements of $ G^{\prime } \setminus G $ might be good candidates to lead to these extrinsic parameters. Nonetheless, a most consistent answer to this burning question can be well understood, for example, by remembering that there is an application [\refcite{birk}]
		\begin{equation}
			\left( g^{\prime } , g^{\prime \prime } \right) \ \mapsto \ \varphi ^{\prime } \left( g^{\prime } , g^{\prime \prime } \right) = g \label{general-group-multiplication}
		\end{equation}
		that gives a group structure to $ G^{\prime } $. After all, just as we were able to define $ U_{f} $ by using (\ref{group-multiplication}), we are also able to define another application
		\begin{equation}
			M_{f} \left( g_{1} , \ldots , g_{k} \right) = \varphi ^{\prime } \left( \ldots \varphi ^{\prime } \left( \varphi ^{\prime } \left( g_{1} , g_{2} \right) , g _{3} \right) , \ldots , g_{k} \right) \label{non-holonomia}
		\end{equation}
		by using (\ref{general-group-multiplication}). And why does (\ref{general-group-multiplication}) help us to understand this most consistent answer? Because as [\refcite{birk}]
		\begin{equation}
			\varphi = \left. \varphi ^{\prime } \right\vert _{G^{\prime } = G} \ , \label{restriction}
		\end{equation}
		this is exactly what allows us to explore the possibility that $ M_{f} $ may be reduced to $ U_{f} $ when, for example, $ G^{\prime } = G $.
		
		Note that, when we defined this new application (\ref{non-holonomia}), we have not made any mention of the possibility of identifying the coordinate $ g_{\ell } $ of the $ k $-tuple $ \left( g_{1} , \ldots , g_{k} \right) $ with the group element assigned to the $ \ell $-th side of a lattice face $ f $. And we did not make any mention of this possibility because, in addition to it being possible for $ G^{\prime } $ to differ from $ G $, we also want to interpret $ M_{f} $ as the application that calculates all $ \left\vert G^{\prime } \right\vert ^{k} $ possible values that can be obtained by multiplying $ k $ arbitrary elements of $ G^{\prime } $. However, although $ M_{f} $ has no commitment to calculate the holonomies of the $ f $-th face of $ \mathcal{L} _{n} $, it is important to observe that it accidentally calculates them: this happens when each of the $ \ell $-th coordinates of $ \left( g_{1} , \ldots , g_{k} \right) $ is the $ \ell $-th element of the sequence $ \mathfrak{g} _{1} , \ldots , \mathfrak{g} _{k} $ of gauge fields that is assigned to this lattice face. And why is this important to observe? Because it is precisely this accidental calculation that allows us to conclude that (\ref{pre-part-function}) can be recovered, for instance, from a more general partition function
		\begin{equation}
			\tilde{Z} = \sum _{\left\{ g \right\} } \prod _{f \in \mathcal{L} _{n}} e^{- \beta \left[ \psi \left( M_{f} \right) + \psi \bigl( M^{-1} _{f} \bigr) \right] } \cdot e^{\lambda _{f} \Phi _{f}} \ , \label{total-part-function}
		\end{equation}
		where $ \lambda _{f} $ is a positive real number and
		\begin{equation}
			\Phi _{f} = \Phi \bigl( \psi \left( M_{f} \right) \bigr) = \ln \delta \left( \psi \left( M_{f} \right) , \psi \left( U_{f} \right) \right) \ . \label{ln-constraint}
		\end{equation}
		After all, as $ \delta \left( \psi \left( M_{f} \right) , \psi \left( U_{f} \right) \right) $ should be interpreted as a Kronecker delta that was written differently only for the sake of intelligibility (i.e., $ \delta \left( \psi \left( M_{f} \right) , \psi \left( U_{f} \right) \right) = \delta _{ab} $, where $ a = \psi \left( M_{f} \right) $ and $ b = \psi \left( U_{f} \right) $), it becomes quite clear that (\ref{total-part-function}) actually leads us to a Hamiltonian formulation \label{discussion}
		\begin{equation}
			\tilde{H} = - \ln \tilde{Z} = \beta \sum _{f \in \mathcal{L} _{n}} \left[ \psi \left( M_{f} \right) + \psi \bigl( M^{-1} _{f} \bigr) \right] + \sum _{f \in \mathcal{L} _{n}} \lambda _{f} \hspace*{0.04cm} \Phi _{f} \label{lattice-hamiltonian-with-constraints}
		\end{equation}
		with constraints, which shows us that
		\begin{equation*}
			\left. \tilde{H} \right\vert _{\Phi = 0} = \beta \sum _{f \in \mathcal{L} _{n}} \left[ \psi \left( U_{f} \right) + \psi \bigl( U^{-1} _{f} \bigr) \right] = - \ln Z = H \ .
		\end{equation*}
		
		\subsection{What else can be said about $ M_{f} $?}
		
			In view of the last results/comments, perhaps you, the reader, are feeling a little uncomfortable. After all, as much as everything we have done seems to be mathematically correct, the use of this application $ M_{f} $ (which, a priori, should not be interpreted as a \textquotedblleft holonomy meter\textquotedblright ) to obtain a Hamiltonian formulation with constraints sounds a bit unnatural, is it not? But, as unnatural as it sounds, there is no way to say that it is wrong to interpret $ M_{f} \left( g_{1} , \ldots , g_{k} \right) $ as a holonomy that is not necessarily physical. And what do we mean \textquotedblleft a holonomy that is not necessarily physical\textquotedblright ?
			
			In order to understand the answer to this question, we should remember that $ U_{f} $ can be associated with a local estimate of how curved is $ \mathcal{M} _{n} $ from the point of view of the faces of $ \mathcal{L} _{n} $ [\refcite{clarke}]. After all, note that, due to the way that $ U_{f} $ was defined in (\ref{holonomia}), it is not difficult to conclude, for instance, that each of the elements of $ G $ can characterize different face deformations of $ \mathcal{L} _{n} $ [\refcite{gambini}]. Observe that this is a conclusion that, due to this same definition (\ref{holonomia}), does not extend to the elements of $ G^{\prime } \setminus G $ when $ G^{\prime } $ is different from $ G $. However, it is important to note that there is nothing to prevent the existence of several hypothetical lattices, which may have the same dimension as $ \mathcal{L} _{n} $, whose face deformations can be characterized by all the elements of $ G^{\prime } $. And why is this important to note? Because, since $ G $ is a subset of $ G^{\prime } $, it is not difficult to conclude that one of these hypothetical lattices is precisely the same $ \mathcal{L} _{n} $ where the physics of our system is defined. Thus, by noting that
			\begin{itemize}
				\item it is not absurd to think that there is another application that can measure all the holonomies of these hypothetical lattices, and
				\item this other application needs to define $ U_{f} $ as a special case by using (\ref{restriction}),
			\end{itemize}
			it is reasonable to identify this other application with the same $ M_{f} $ that we already defined in (\ref{non-holonomia}).
			
		\subsection{\label{first-class} The first-class constraints}
		
			Once the interpretation of $ M_{f} $ is already well understood, it is also worth mentioning that the result
			\begin{equation*}
				\left. \sum _{\left\{ g \right\}} \prod _{f \in \mathcal{L} _{n}} e^{- \beta \left[ \psi \left( M_{f} \right) + \psi \bigl( M^{-1} _{f} \bigr) \right] } \cdot e^{\lambda _{f} \hspace*{0.04cm} \Phi _{f} } \right\vert _{\Phi _{f} = 0} = \sum _{\left\{ \mathfrak{g} \right\}} \prod _{f \in \mathcal{L} _{n}} e^{- \beta \left[ \psi \left( U_{f} \right) + \psi \bigl( U^{-1} _{f} \bigr) \right] } \ ,
			\end{equation*}
			which allows us to get (\ref{pre-part-function}) as
			\begin{equation*}
				Z = \left. \tilde{Z} \hspace*{0.04cm} \right\vert _{\Phi = 0} \ ,
			\end{equation*}			
			can be interpreted in terms of conditional probabilities [\refcite{kolmogorov}]. After all, by considering that $ P_{f} \left( A \right) = e^{- \beta \left[ \psi \left( M_{f} \right) + \psi \bigl( M^{-1} _{f} \bigr) \right] } $ is the probability of the $ f $-th face holonomy to be equal to $ M_{f} $ (event $ A $), and $ P_{f} \left( B \right) = \left. e^{\lambda _{f} \hspace*{0.04cm} \Phi _{f} } \right\vert _{\Phi _{f} = 0} $ is the probability of $ \psi \left( M_{f} \right) = \psi \left( U_{f} \right) $ occurring (event $ B $), it is not difficult to conclude that the probability $ P_{f} \left( A \cap B \right) = e^{- \beta \left[ \psi \left( U_{f} \right) + \psi \bigl( U^{-1} _{f} \bigr) \right] } $ of the $ f $-th face holonomy to be equal to $ U_{f} $ (event $ A \cap B $) satisfies
			\begin{equation*}
				\sum _{\left\{ g \right\}} P_{f} \left( A \cap B \right) = \sum _{\left\{ \mathfrak{g} \right\}} P_{f} \left( A \hspace*{0.04cm} \vert \hspace*{0.04cm} B \right) \cdot P_{f} \left( B \right)
			\end{equation*}
			since the events $ A $ and $ B $ are independent (i.e., $ P_{f} \left( A \hspace*{0.04cm} \vert \hspace*{0.04cm} B \right) = P_{f} \left( A \right) $). Here, the sum variable $ \left\{ g \right\} $ is analogous to $ \left\{ \mathfrak{g} \right\} $: i.e., the sum variable $ \left\{ g \right\} $ in (\ref{total-part-function}) means that all the possible configurations of the elements of $ G^{\prime } $ are being computed in $ Z $. In this way, although it seems that we choose to define $ \tilde{Z} $ as (\ref{total-part-function}) just because it results in a Hamiltonian formulation with constraints, it is valid to say that our choice was also based on probabilistic considerations.
			
			Anyway, given this realization of the pure gauge lattice systems as Hamiltonian systems with constraints, it is important to conclude this Section by showing that $ \Phi _{f} $ actually can be interpreted as the discretized version of the first-class constraints. And one of the ways to show this is to observe that nothing changes if, for instance, we use
			\begin{equation*}
				\Phi _{\alpha ,f} = \Phi _{\alpha } \bigl( \psi \left( M_{f} \right) \bigr) = \ln \left[ \delta \left( \psi \left( M_{f} \right) , \psi \left( U_{f} \right) \right) ^{\alpha } \right] \ ,
			\end{equation*} % \label{general-function}
			where $ \alpha $ is a real number such that $ 0 < \alpha < \infty $, instead of using (\ref{ln-constraint}) because
			\begin{eqnarray*}
				\lefteqn{\Phi _{f} = \ln \delta \left( \psi \left( M_{f} \right) , \psi \left( U_{f} \right) \right) = 0} \hspace*{1.0cm} \\
				& \Rightarrow & \Phi _{\alpha , f} = \ln \left[ \delta \left( \psi \left( M_{f} \right) , \psi \left( U_{f} \right) \right) ^{\alpha } \right] = \alpha \hspace*{0.04cm} \ln \delta \left( \psi \left( M_{f} \right) , \psi \left( U_{f} \right) \right) = 0 \ .
			\end{eqnarray*}
			And since this allows us to see, for example, that the last term in (\ref{lattice-hamiltonian-with-constraints}) is such that
			\begin{equation*}
				\sum _{f \in \mathcal{L} _{n}} \lambda _{f} \hspace*{0.04cm} \ln \delta \left( \psi \left( M_{f} \right) , \psi \left( U_{f} \right) \right) = 0 \ \Rightarrow \ \sum _{f \in \mathcal{L} _{n}} \left( \lambda _{f} \cdot \alpha \right) \hspace*{0.04cm} \ln \delta \left( \psi \left( M_{f} \right) , \psi \left( U_{f} \right) \right) = 0 \ ,
			\end{equation*}
			it also becomes clear that the Lagrange multipliers $ \lambda _{f} $, which implements the constraints $ \Phi _{f} = 0 $ in (\ref{lattice-hamiltonian-with-constraints}), cannot be unequivocally determined. In other words, at the same time that these last three expressions shows us that there are infinite choices that can be made for the constraints that define $ \mathcal{L} _{n} $, these same expressions also make it clear that, for any constraint, the Lagrange multiplier is not uniquely determined. And as 
			\begin{itemize}
				\item the elements of $ G^{\prime } $ lead us to an extrinsic parameterization of $ \mathcal{L} _{n} $ through of $ \psi $, when $ \psi \left( M_{f} \right) \neq \psi \left( U_{f} \right) $, and (therefore)
				\item this resonates with what was said in the Introduction about the pair $ \Omega = \left( Q , P \right) $ that parameterizes $ \left( T^{\ast } \mathcal{M} _{n} \right) ^{\perp } $ intrinsically (i.e., that is extrinsic to $ T^{\ast } \mathcal{M} _{n} $),
			\end{itemize}
			the interpretation of these finite-group gauge theories on lattices as Hamiltonian systems with constraints becomes quite evident.
			
			While this last observation is already enough for us to indirectly conclude that all equations $ \Phi _{\alpha ,f} = 0 $, which can be defined with the infinite values of $ \alpha $ \footnote{Note that $ \Phi _{f} = \Phi _{1,f} $.}, can be interpreted as first-class constraints, it is interesting to note that this same conclusion can also be obtained directly from the calculation of
			\begin{equation}
				\left[ \Phi _{\alpha ,f} , \Phi _{\alpha ^{\prime } ,f^{\prime }} \right] \ . \label{dirac-parenthesis}
			\end{equation}
			After all, according to the (quantum version of the) Dirac’s consistency conditions [\refcite{dirac,david}]
			\begin{eqnarray*}
				\dot{\Phi } _{\alpha ,f} & = & \bigl[ \Phi _{\alpha ,f} , \tilde{H} \bigr] \\
				& = & \beta \sum _{f^{\prime } \in \mathcal{L} _{n}} \bigl[ \Phi _{\alpha ,f} , \psi \left( M_{f} \right) + \psi \bigl( M^{-1} _{f} \bigr) \bigr] + \sum _{f^{\prime } \in \mathcal{L} _{n}} \lambda _{f^{\prime }} \left[ \Phi _{\alpha ,f} , \Phi _{\alpha ^{\prime } ,f^{\prime }} \right] = 0 \ ,
			\end{eqnarray*}
			if all these equations $ \Phi _{\alpha ,f} = 0 $ can, indeed, be interpreted as first-class constraints, (\ref{dirac-parenthesis}) needs to vanish on $ \mathcal{L} _{n} $ for all values of $ \alpha $, $ \alpha ^{\prime } $, $ f $ and $ f^{\prime } $. And by taking into account that
			\begin{equation*}
				\left( \ln x \right) \left( \ln y \right) = \left( \ln y \right) \left( \ln x \right) \ \Leftrightarrow \ \underbrace{\left( \ln x \right) \left( \ln y \right) - \left( \ln y \right) \left( \ln x \right) } _{\left[ \ln x , \ln y \right] } = 0
			\end{equation*}
			holds for any non-negative real numbers $ x $ and $ y $, it is not difficult to conclude that this is exactly what happens.
			
			In view of what was demonstrated in the last paragraph, there is no denying that the fact that all the functions $ \Phi _{\alpha ,f} $ commute between them further reinforces, now from another point of view, that all the Lagrange multipliers in (\ref{lattice-hamiltonian-with-constraints}) really cannot be unequivocally determined. But, for the sake of completeness, it is also of paramount importance to end this Section keeping in mind that, even though this last demonstration was done in a very simple way, $ \Phi _{\alpha ,f} $ is a function of a class function. And why is it of paramount importance to end this Section with this in mind? Because every function of a class function can be also interpreted as a class function [\refcite{isaacs}]. After all, as this interpretation also extends to the sum $ \psi \left( M_{f} \right) + \psi \bigl( M^{-1} _{f} \bigr) $, we can conclude that all these Lagrange multipliers are also class functions. That is, even though all these Lagrange multipliers are real numbers that cannot be unequivocally determined, all of them are also class functions of the elements of $ G^{\prime } $ that, as noted earlier, lead us to an extrinsic parameterization of $ \mathcal{L} _{n} $. %%%%%% 83B
	
	\section{\label{qdm}The Kitaev Quantum Double Models as an example}
	
		Before we conclude this paper, it is interesting to cite an example of lattice gauge theories, where its interpretation as a Hamiltonian system with constraints is already quite clear. After all, in addition to this example being useful for those who are being introduced to the concept of (lattice) gauge theory, it is also useful for those who, because they already have some familiarity with models that support some kind of quantum computing, may be under the impression that they have seen it all (what we have discussed here) somewhere before. And what useful example is this? This is the class of the \emph{Kitaev Quantum Double Models} ($ D \left( G \right) $), so named in honour of Alexei Yu. Kitaev and because they satisfy the \emph{Drinfeld's quantum double algebra} [\refcite{kitaev-1,drinfeld}].
		
		There are several references that can be used to understand the various features of these $ D \left( G \right) $ models [\refcite{pachos,mf-pedagogical,brell,aguado,miguel-3d,naaijkens,miguel-deformed,bullivant,komar,cha,wang-h,cui}] and, precisely because of that, we will not use this Section to detail them. But something that should be said about the $ D \left( G \right) $ models is that, for instance, they were defined to be deliberately interpreted as lattice gauge theories. And two reasons that support this assertion are:
		\begin{itemize}
			\item[(i)] classical and quantum computing need to be done using/manipulating real objects (i.e., physical objects) which, therefore, obey the laws of Physics; and
			\item[(ii)] the quantum theories that describe electromagnetic phenomena (i.e., the phenomena that occur in any material that can be used to perform any of these computations) are gauge theories.
		\end{itemize}
		
		But, before we explain why these $ D \left( G \right) $ models are examples of lattice gauge theories, it is very important to mention that the quantum foundation of these computational models is based on the manipulations of \emph{quantum dits} (qudits): i.e., on the manipulations of a quantum version of the $ d $-ary digits, which can be described as a unitary vector of a $ d $-dimensional Hilbert space $ \mathfrak{H} _{d} $ [\refcite{bryl,wang-y}]. And given this computational context, it is also very important to mention that, in order to avoid any problems with reading the data encoded by these qudits, these $ D \left( G \right) $ models are defined by associating one $ \mathfrak{H} _{d} $ with each of the edges of an oriented lattice $ \mathcal{L} _{2} $, which discretizes a $ 2 $-dimensional compact orientable (sub)manifold $ \mathcal{M} _{2} $. In other words, as with the lattice gauge theories evaluated in this paper, it is already clear that these $ D \left( G \right) $ models are, for instance, deliberately defined by using an oriented lattice $ \mathcal{L} _{2} $ in which a vector is allocated on each of its edges.
				
		\subsection{Why can these $ D \left( G \right) $ models be interpreted as Hamiltonian systems with constraints?}
		
			Of course, the fact that the $ D \left( G \right) $ models are defined by allocating vectors of a Hilbert space to edges of $ \mathcal{L} _{2} $ is not enough to characterize these models as lattice gauge theories. But, by noting that these $ D \left( G \right) $ models are defined by taking $ \mathcal{B} = \big\{ \left\vert \mathfrak{g} \right\rangle : \mathfrak{g} \in G \big\} $ as the single-qudit computational basis of $ \mathfrak{H} _{d} $ [\refcite{kitaev-1,mf-pedagogical}], this \textquotedblleft paves the way\textquotedblright \hspace*{0.01cm} for such a characterization. After all, since the vectors (kets) of this orthonormal basis are indexed by the elements of a group $ G $, this allows us to define, for instance, two operations
			\begin{equation}
				L^{\left( \mathfrak{g} \right) } _{+} \left\vert \mathfrak{g} ^{\prime } \right\rangle = \left\vert \mathfrak{g} \cdot \mathfrak{g} ^{\prime } \right\rangle \quad \textnormal{and} \quad L^{\left( \mathfrak{g} \right) } _{-} \left\vert \mathfrak{g} ^{\prime } \right\rangle = \left\vert \mathfrak{g} ^{\prime } \cdot \mathfrak{g} ^{-1} \right\rangle \label{l-plus-minus-operations}
			\end{equation}
			by using the same multiplications that have already been mentioned in the items (a) and (b) on page \pageref{gauge-transformation}. That is, it opens up the possibility of defining a single operator $ A^{\left( \mathfrak{g} \right) } _{v} $ that, when acting on the edge subset $ S_{v} $ that gives structure to the $ v $-th vertex, performs a transformation similar to that presented in Subsection \ref{gauge-subsection}: i.e., a single operator given by
			\begin{equation*}
				A^{\left( \mathfrak{g} \right) } _{v} = \prod _{\ell \in S_{v}} L^{\left( \mathfrak{g} \right) } _{\ell } \ ,
			\end{equation*}
			where $ L^{\left( \mathfrak{g} \right) } _{\ell } $ acts as
			\begin{itemize}
				\item[(a')] $ L^{\left( \mathfrak{g} \right) } _{+} $, if the $ \ell $-th edge orientation pointing out of this $ v $-th vertex, or
				\item[(b')] $ L^{\left( \mathfrak{g} \right) } _{-} $, otherwise.
			\end{itemize}
			
			Since we just talked about this operator $ A^{\left( \mathfrak{g} \right) } _{v} $, a natural question that you, the reader, may be asking right now is: how does the possibility of defining this $ A^{\left( \mathfrak{g} \right) } _{v} $ help us, for instance, to interpret these $ D \left( G \right) $ models as lattice gauge theories? By noting that the $ D \left( G \right) $ Hamiltonian operator [\refcite{kitaev-1}]
			\begin{equation}
				H_{D \left( G \right) } = \sum _{v \in \mathcal{L} _{2}} \left( \mathds{1} _{v} - A_{v} \right) + \sum _{s \in \mathcal{L} _{2}} \left( \mathds{1} _{s} - B_{s} \right) \ , \label{dg-hamiltonian}
			\end{equation}
			which can be obtained by taking the logarithm of a partition function as explained in Ref. [\refcite{miguel-3d}], is defined by using a vertex operator given by [\refcite{kitaev-2}]
			\begin{equation*}
				A_{v} = \frac{1}{\left\vert G \right\vert } \sum _{\mathfrak{g} \in G} A^{\left( \mathfrak{g} \right) } _{v} \ .
			\end{equation*}
			After all, this operator $ A_{v} $, which (unlike $ \mathds{1} _{v} $) acts not identically only on $ S_{v} $, averages out the possible transformations that $ A^{\left( \mathfrak{g} \right) } _{v} $ is able to do by using all elements of $ G $.
			
			\subsubsection{How does the operator $ B_{s} $ act on $ \mathcal{L} _{2} $?}
			
				Given what we have seen so far, it is tempting to conclude that $ A^{\left( \mathfrak{g} \right) } _{v} $ performs lattice gauge transformations. But before confirming this conclusion, it is important to identify an operator that is capable of measuring the lattice face holonomies in these $ D \left( G \right) $ models. And while we still have not said anything about the operator $ B_{s} $ that appears in (\ref{dg-hamiltonian}), it is interesting to point out that this is exactly what $ B_{s} $ does when acting on the lattice site $ s = \left( v , f \right) $.
			
\begin{figure}[!t]
					\begin{center}
						\begin{tikzpicture}[
							scale=0.3,
							equation/.style={thin},
							trans/.style={color=gray, ultra thick},
							function/.style={->, color=gray, ultra thick, >=stealth}
							]
							\draw[function,color=gray] (24,-2) -- (20,-1.0) node[above=2pt,right=-1pt] {};
							\draw[function,color=gray] (16,0) -- (19.0,4) node[above=2pt,right=-1pt] {};
							\draw[function,color=gray]  (24,-2) -- (26.6,1.1) node[above=2pt,right=-1pt] {};
							\draw[function,color=gray] (22,8) -- (25.5,6) node[above=2pt,right=-1pt] {};
							\draw[trans,color=gray] (16,0) -- (14,-0.5) node[above=2pt,right=-1pt] {};
							\draw[trans,color=gray] (24,-2) -- (25,-3) node[above=2pt,right=-1pt] {};
							\draw[trans,color=gray] (22,8) -- (24,9) node[above=2pt,right=-1pt] {};
							\draw[trans,color=gray] (22,8) -- (19.5,9) node[above=2pt,right=-1pt] {};
							\draw[trans,color=gray] (29,4) -- (30,4) node[above=2pt,right=-1pt] {};	
							\draw[trans,color=gray] (16,0) -- (24,-2) node[above=2pt,right=-1pt] {};
							\draw[trans,color=gray] (16,0) -- (22,8) node[above=2pt,right=-1pt] {};
							\draw[trans,color=gray] (22,8) -- (29,4) node[above=2pt,right=-1pt] {};
							\draw[trans,color=gray] (26.5,1.0) -- (29.0,4) node[above=2pt,right=-1pt] {};
							\draw [rotate=15,fill=yellow!50,thin,line width=0.7pt] (21.9,-5.8) arc[x radius=0.8cm, y radius =3.3cm, start angle=-180, end angle=180];
							\draw[color=black,fill=blue] (22.8,2.6) circle (1.5ex);
							\draw[color=gray,fill=gray] (16,0) circle (1.5ex);
							\draw[color=black,fill=red] (24,-2) circle (1.5ex);
							\draw[color=gray,fill=gray] (22,8) circle (1.5ex);
							\draw[color=gray,fill=gray] (29,4) circle (1.5ex);
						\end{tikzpicture} \hspace*{1.0cm}
						\begin{tikzpicture}[
							scale=0.3,
							equation/.style={thin},
							trans/.style={color=gray, ultra thick},
							function/.style={->, color=gray, ultra thick, >=stealth}
							]
							\draw [color=olive,thin,->,line width=1.0pt,>=stealth] (24.5,3) arc[x radius=1.8cm, y radius =1.8cm, start angle=0, end angle=330];
							\draw[function,color=gray] (24,-2) -- (20,-1.0) node[above=2pt,right=-1pt] {};
							\draw[function,color=gray] (16,0) -- (19.0,4) node[above=2pt,right=-1pt] {};
							\draw[function,color=gray]  (24,-2) -- (26.6,1.1) node[above=2pt,right=-1pt] {};
							\draw[function,color=gray] (22,8) -- (25.5,6) node[above=2pt,right=-1pt] {};
							\draw[trans,color=gray] (16,0) -- (14,-0.5) node[above=2pt,right=-1pt] {};
							\draw[trans,color=gray] (24,-2) -- (25,-3) node[above=2pt,right=-1pt] {};
							\draw[trans,color=gray] (22,8) -- (24,9) node[above=2pt,right=-1pt] {};
							\draw[trans,color=gray] (22,8) -- (19.5,9) node[above=2pt,right=-1pt] {};
							\draw[trans,color=gray] (29,4) -- (30,4) node[above=2pt,right=-1pt] {};	
							\draw[trans,color=gray] (16,0) -- (24,-2) node[above=2pt,right=-1pt] {};
							\draw[trans,color=gray] (16,0) -- (22,8) node[above=2pt,right=-1pt] {};
							\draw[trans,color=gray] (22,8) -- (29,4) node[above=2pt,right=-1pt] {};
							\draw[trans,color=gray] (26.5,1.0) -- (29.0,4) node[above=2pt,right=-1pt] {};
							\draw[equation] (26.9,0.4) -- (26.9,0.4) node[above=2pt,right=-1pt] {$ \left\vert \mathfrak{g} _{1} \right\rangle $};
							\draw[equation] (24.8,7.2) -- (24.8,7.2) node[above=2pt,right=-1pt] {$ \left\vert \mathfrak{g} _{2} \right\rangle $};
							\draw[equation] (15.8,4.0) -- (15.8,4.0) node[above=2pt,right=-1pt] {$ \left\vert \mathfrak{g} _{3} \right\rangle $};
							\draw[equation] (19.1,-2.7) -- (19.1,-2.7) node[above=2pt,right=-1pt] {$ \left\vert \mathfrak{g} _{4} \right\rangle $};
							\draw[color=gray,fill=gray] (16,0) circle (1.5ex);
							\draw[color=black,fill=red] (24,-2) circle (1.5ex);
							\draw[color=gray,fill=gray] (22,8) circle (1.5ex);
							\draw[color=gray,fill=gray] (29,4) circle (1.5ex);
						\end{tikzpicture}
					\end{center}
					\caption{\protect\label{holonomy-site} On the left, we see (in baby yellow colour) an example of lattice site: i.e., an ordered pair $ s = \left( v , f \right) $, which is composed of one vertex $ v $ and one face $ f $ (highlighted in red and blue colours respectively), that helps to define how the operators $ B^{\left( \mathfrak{g} \right) } _{s} $ act on the faces of $ \mathcal{L} _{2} $. After all, even though each lattice face is indexed by a single value of $ f $, the way these operators are defined in the $ D \left( G \right) $ models requires, for instance, that there is some kind of \textquotedblleft origin\textquotedblright \hspace*{0.01cm} from which the holonomy of the $ f $-th face can be estimated. And, according to what we see on the right, this \textquotedblleft origin\textquotedblright \hspace*{0.01cm} is precisely the vertex $ v $ that defines the site $ s $, since this holonomy $ U^{\left( v \right) }_{f} $ is calculated by taking, as the first term of this calculation, the group element that is associated with the first ket/edge appearing in counterclockwise order. That is, in the case of this lattice face that we see on the right, this calculation is done as $ U^{\left( v \right) } _{f} = \left( \mathfrak{g} _{1} \right) ^{-1} \cdot \mathfrak{g} _{2} \cdot \left( \mathfrak{g} _{3} \right) ^{-1} \cdot \left( \mathfrak{g} _{4} \right) ^{-1} $ in deference, for instance, to what has already been explained in Figures \ref{clockwise-holonomy} and \ref{counterclockwise-holonomy}.}
				\end{figure}
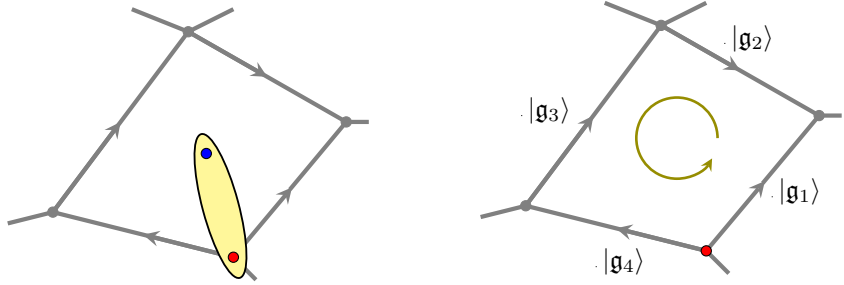

				In order to understand how $ B_{s} $ does this, it is relevant to note, for instance, that the operations mentioned in (\ref{l-plus-minus-operations}) are not the only ones that we can define by using $ \mathcal{B} $: two others are
				\begin{equation}
					T^{\left( \mathfrak{g} \right) } _{+} \left\vert \mathfrak{g} ^{\prime } \right\rangle = \delta \left( \mathfrak{g} , \mathfrak{g} ^{\prime } \right) \left\vert \mathfrak{g} ^{\prime } \right\rangle \quad \textnormal{and} \quad T^{\left( \mathfrak{g} \right) } _{-} \left\vert \mathfrak{g} ^{\prime } \right\rangle = \delta \left( \mathfrak{g} ^{-1} , \mathfrak{g} ^{\prime } \right) \left\vert \mathfrak{g} ^{\prime } \right\rangle \ . \label{t-plus-minus-operations}
				\end{equation}
				And since $ \delta \left( a , b \right) = \delta _{ab} $ is a Kronecker delta, another operator that we can define with the help of these operations (\ref{t-plus-minus-operations}) is [\refcite{kitaev-1}]
				\begin{equation}
					B^{\left( \mathfrak{g} \right) } _{s} = \sum _{U^{\left( v \right) }_{f} = \mathfrak{g}} \left( \prod _{\ell \in S_{f}} T^{\left( \mathfrak{g} \right) } _{\ell } \right) \ , \label{b-definition}
				\end{equation}			
				which acts on the edge subset $ S_{f} $ that gives structure to the $ f $-th face, by using an operator $ T^{\left( \mathfrak{g} \right) } _{\ell } $ that acts as
				\begin{itemize}
					\item $ T^{\left( \mathfrak{g} \right) } _{+} $, if the $ \ell $-th edge is oriented counterclockwise from the point of view of the $ f $-th face, or
					\item $ T^{\left( \mathfrak{g} \right) } _{-} $, otherwise.
				\end{itemize}
				Here, $ U^{\left( v \right) }_{f} $ calculates the $ f $-th face holonomy (i) by using the counterclockwise direction and (ii) by taking, as the first term of its product, the group element that is associated with one of the edges that is delimited by the $ v $-th vertex. After all, as the operators $ B^{\left( \mathfrak{g} \right) } _{s} $ are defined without using any class function explicitly (and, therefore, the possibility that different holonomies characterize the same lattice face is non-zero), so that the \textquotedblleft scan\textquotedblright \hspace*{0.01cm} performed by the operator $ B_{s} $ in (\ref{dg-hamiltonian}) to lead to consistent results, the operators $ B^{\left( \mathfrak{g} \right) } _{s} $ are deliberately defined by using a sum restricted to $ U^{\left( v \right) }_{f} = \mathfrak{g} $. This calculation is well illustrated in Figure \ref{holonomy-site}.
				
				Note that, as artificial as $ B^{\left( \mathfrak{g} \right) } _{s} $ may seem, its definition makes it possible to interpret it, for instance, as a projector and that is precisely what allows us to recognize it as a kind of \textquotedblleft holonomy meter\textquotedblright . After all, since the sum in (\ref{b-definition}) is constrained to the fact that $ U^{\left( v \right) }_{f} $ must be equal to $ \mathfrak{g} $, the Kronecker deltas in (\ref{t-plus-minus-operations}) make all the measurements, which are performed by $ B^{\left( \mathfrak{g} \right) } _{s} $ on any (site that characterizes a) face of $ \mathcal{L} _{2} $, always equal to
				\begin{itemize}
					\item $ 1 $, if $ U^{\left( v \right) }_{f} = \mathfrak{g} $, and
					\item $ 0 $, otherwise.
				\end{itemize}
				But, despite what we are saying is correct, the operator that appears in (\ref{dg-hamiltonian}) is $ B_{s} $ and not $ B^{\left( \mathfrak{g} \right) } _{s} $. Hence, the question that needs to be answered now is: what is the definition of $ B_{s} $ and what does it have to do with $ B^{\left( \mathfrak{g} \right) } _{s} $? And since we already stated that $ B_{s} $ is capable of measuring the lattice face holonomies, the answer to this question is [\refcite{kitaev-1,kitaev-2}]:
				\begin{equation}
					B_{s} \equiv B^{\left( \mathfrak{e} \right) } _{s} \ , \label{b-operators}
				\end{equation}
				where $ \mathfrak{e} $ is the neutral element of $ G $. That is, $ B_{s} $ is a special case of the operators $ B^{\left( \mathfrak{g} \right) } _{s} $ that measures flat connections (i.e., trivial holonomies that are characterized by $ \mathfrak{h} = \mathfrak{e} $ along the faces) in the $ D \left( G \right) $ models.
				
				Given this scenario, it is finally possible to confirm the interpretation of $ A^{\left( \mathfrak{g} \right) } _{v} $ as an operator that performs lattice gauge transformations. And for this to be done, it is crucial to note that, since
				\begin{eqnarray*}
					L^{\left( \mathfrak{g} \right) } _{\pm } L^{\left( \mathfrak{h} \right) } _{\pm } = L^{\left( \mathfrak{h} \right) } _{\pm } L^{\left( \mathfrak{g} \right) } _{\pm } & , & L^{\left( \mathfrak{g} \right) } _{\pm } L^{\left( \mathfrak{h} \right) } _{\mp } = L^{\left( \mathfrak{h} \right) } _{\mp } L^{\left( \mathfrak{g} \right) } _{\pm } \ , \\
					L^{\left( \mathfrak{g} \right) } _{\pm } T^{\left( \mathfrak{h} \right) } _{\pm } = T^{\left( \mathfrak{gh} \right) } _{\pm } L^{\left( \mathfrak{g} \right) } _{\pm } & , & L^{\left( \mathfrak{g} \right) } _{\pm } T^{\left( \mathfrak{h} \right) } _{\mp } = T^{\left( \mathfrak{hg} ^{-1} \right) } _{\mp } L^{\left( \mathfrak{g} \right) } _{\pm } \ , \\
					T^{\left( \mathfrak{g} \right) } _{\pm } T^{\left( \mathfrak{h} \right) } _{\pm } = T^{\left( \mathfrak{h} \right) } _{\pm } T^{\left( \mathfrak{g} \right) } _{\pm } \quad & \textnormal{and} & \quad T^{\left( \mathfrak{g} \right) } _{\pm } T^{\left( \mathfrak{h} \right) } _{\mp } = T^{\left( \mathfrak{h} \right) } _{\mp } T^{\left( \mathfrak{g} \right) } _{\pm } \ ,
				\end{eqnarray*}
				$ A_{v} $ and $ B^{\left( \mathfrak{g} \right) } _{s} $ commute among them for all values of $ v $ and $ s $. After all, since these operators commute among them, this means that all the holonomies measured by $ B^{\left( \mathfrak{g} \right) } _{s} $ are not changed by the action of $ A_{v} $ on $ \mathcal{L} _{2} $. In other words, and in light of what was discussed in Section \ref{lattice}, $ A_{v} $ actually performs lattice gauge transformations because its action on $ \mathcal{L} _{2} $ does not modify the face holonomies [\refcite{pachos}]. 
			
			\subsubsection{Some geometric considerations on these $ D \left( G \right) $ models.}
				
				Just for the sake of completeness, it is important to point out here that, as with all the operators $ B^{\left( \mathfrak{g} \right) } _{s} $, the vertex operator $ A_{v} $ can also be interpreted as a projector [\refcite{kitaev-1,pachos}]. After all, as much as it actually replaces the kets associated with $ S_{v} $ for others, it is not difficult to prove that its action, on all the vertices of $ \mathcal{L} _{2} $, does not change the encoding written in this lattice. That is, as with all the operators $ B^{\left( \mathfrak{g} \right) } _{s} $, this vertex operator $ A_{v} $ is incapable of modifying the state of this lattice system and its eigenvalues are equal to $ 0 $ and $ 1 $.
				
				As a matter of fact, as a consequence of the Hamiltonian (\ref{dg-hamiltonian}) being nothing less than a superposition of several operators $ A_{v} $ and $ B_{s} $, it is not wrong to say that it is precisely this fact, which $ A_{v} $ and $ B_{s} $ are projectors, that makes $ H_{D \left( G \right) } $ able to measure the energies of these $ D \left( G \right) $ models. And in order to understand how $ H_{D \left( G \right) } $ measures these energies, it is interesting to note that, since $ \mathds{1} _{s} $ is the identity operator that acts effectively on $ S_{f} $, the smallest eigenvalue of $ H_{D \left( G \right) } $ is $ 0 $. After all, as $ 0 $ and $ 1 $ are the only values that $ A_{v} $ and $ B_{s} $ can measure by acting on $ \mathcal{L} _{2} $, what these operators do, from the $ H_{D \left( G \right) } $ point of view, can be interpreted as a count of how many local elevations of energy (i.e., of how many \emph{quasiparticles}) there are on this lattice. Observe that, since the smallest eigenvalue of $ H_{D \left( G \right) } $ is associated with an eigenstate $ \bigl\vert \xi _{0} \bigr\rangle $, where [\refcite{kitaev-1,pachos,kitaev-2}]
				\begin{equation}
					A_{v} \hspace*{0.04cm} \bigl\vert \xi _{0} \bigr\rangle = \bigl\vert \xi _{0} \bigr\rangle \quad \textnormal{and} \quad B_{s} \hspace*{0.04cm} \bigl\vert \xi _{0} \bigr\rangle = \bigl\vert \xi _{0} \bigr\rangle \label{protege}
				\end{equation}
				hold for all the possible values of $ v $ and $ s $, it is not wrong to say, for instance, that these quasiparticles can be created by violating at least one of the conditions in (\ref{protege}).
				
				Nevertheless, since the main purpose of this Section is to show why these $ D \left( G \right) $ models, which are good examples of finite-group gauge theories on lattices, can be interpreted as Hamiltonian systems with constraints, it is interesting to analyse (\ref{protege}) by taking a geometric point of view. After all, as (\ref{b-operators}) tells us that the smallest eigenvalue of $ H_{D \left( G \right) } $ is obtained only when, for instance, $ U^{\left( v \right) }_{f} = \mathfrak{e} $ holds for all the values of $ s $, it is not difficult to conclude that, when these $ D \left( G \right) $ models are in their ground states, $ \mathcal{L} _{2} $ is locally flat. And since $ \mathcal{L} _{2} $ discretizes a $ 2 $-dimensional compact orientable (sub)manifold $ \mathcal{M} _{2} $, this conclusion makes sense because, as in the $ D \left( G \right) $ ground states there are no quasiparticles, each of the lattice faces can be seen as locally flat since they are rough approximations of some open sets of $ \mathcal{M} _{2} $.
				
				Note that, as the quasiparticles detectable by $ B_{s} $ are always such that $ U^{\left( v \right) }_{f} \neq \mathfrak{e} $, the geometric point of view associated with this condition allows us to say, for instance, that the presence/existence of these quasiparticles must be interpreted as local deformations of $ \mathcal{L} _{2} $. And bearing in mind that quantum-computational models try/need to model some reality that can be physically implemented, this interpretation is not strange at all because there are several physical theories that, for instance, assign some spatial curvature to the presence of fields/particles. In fact, a good example of these physical theories is the Standard Model of elementary particles (SM) [\refcite{schwartz}] since, due to the presence of several covariant derivatives [\refcite{man-gr,elon-variedades,man-gd}] in its Lagrangian formulation, it is impossible not to recognize that its fermions are responsible for curving space in some sense [\refcite{derdzinski}]. In plain English, this good example is in line with what we said at the beginning of the Section \ref{lattice} because, in addition to the SM being an example of gauge theory [\refcite{ait-hey-1,ait-hey-2}], it was precisely some of its problems, which required a non-perturbative solution, that have fostered the development of the lattice gauge theories [\refcite{rothe,wilson-loops}].
				
				Nevertheless, in view of this geometric scenario, in which we see that quasiparticles detectable by $ B_{s} $ are responsible for deforming $ \mathcal{L} _{2} $ locally, it is worth noting that quasiparticles that are detectable only by $ A_{f} $ are not capable of doing the same thing. In other words, as the operators $ B^{\left( \mathfrak{g} \right) } _{s} $ (and not $ A^{\left( \mathfrak{g} \right) } _{v} $) are interpreted as \textquotedblleft holonomy meters\textquotedblright , $ B_{s} $ is the only operator in (\ref{dg-hamiltonian}) that can check if there is any deformed face in $ \mathcal{L} _{2} $ and, therefore, define the constraints that determine this lattice. After all, observe that, analogously to what we saw on page \pageref{discussion}, the $ D \left( G \right) $ Hamiltonian operator is nothing more than an
				\begin{equation*}
					H_{D \left( G \right) } = \left. \tilde{H} _{D \left( G \right) } \right\vert _{\Phi = 0} \ ,
				\end{equation*}
				where
				\begin{equation*}
					\tilde{H} _{D \left( G \right) } = \sum _{v \in \mathcal{L} _{2}} \left( \mathds{1} _{v} - A_{v} \right) + \sum _{s \in \mathcal{L} _{2}} \left[ \mathds{1} _{s} - B^{\left( \mathfrak{g} \right) } _{s} \right] \quad \textnormal{and} \quad \Phi _{f} = \ln \delta \bigl( U^{\left( v \right) }_{f} , \mathfrak{e} \bigr) \ .
				\end{equation*} %%%%%% 81B
	
	\section{\label{conclusion} Final remarks}
	
		In view of everything we have presented in this review, it is quite clear that the lattice gauge theories, where $ G $ is a finite group, can indeed be interpreted in terms of Hamiltonian systems with constraints, similarly to what happens in the classical (continuous) gauge (field) theories. After all, as unnatural as it may seem to rewrite (\ref{pre-part-function}) using conditional probabilities, the fact is that, as this gauge system is constrained to $ \mathcal{L} _{n} $ (i.e., to the lattice that is defined when $ \Phi _{f} = 0 $ holds for all the values of $ f $), it makes physical sense to get a zero partition function when $ \left. e^{\lambda _{f} \hspace*{0.04cm} \Phi _{f} } \right\vert _{\Phi _{f} \neq 0} $ holds for, at least, one value of $ f $.
		
		Given that the function $ \Phi _{f} $ is just one of the infinite functions $ \Phi _{\alpha ,f} $ that are capable of defining the same $ \mathcal{L} _{n} $ (i.e., given that $ \Phi _{f} = 0 $ is equivalent to $ \Phi _{\alpha ,f} = 0 $ for all $ 0 < \alpha < \infty $), and that this implies that there are infinite choices that we can make for all the Lagrangian multipliers $ \lambda _{f} $, which implement any of the constraints $ \Phi _{\alpha ,f} = 0 $ to (\ref{lattice-hamiltonian-with-constraints}) without ever-changing $ P_{f} \left( B \right) $, it was possible to infer that all these constraints can be interpreted as first-class. And although this inference was made without calculating the quantum version of the Dirac brackets $ \left[ \Phi _{\alpha ,f} , \Phi _{\alpha ^{\prime } ,f^{\prime }} \right] $, this calculation, which was made in the final of the Section \ref{constrained-characterisation}, only endorsed this interpretation because these brackets are equal to zero for all values of $ \alpha $, $ \alpha ^{\prime } $, $ f $ and $ f^{\prime } $. In this fashion, by remembering that
		\begin{itemize}
			\item these constraints $ \Phi _{\alpha ,f} = 0 $ are also invariant under lattice gauge transformations, and
			\item the elements of the finite gauge group lead us to an intrinsic parameterization of $ \mathcal{L} _{n} $ through of $ \psi $,
		\end{itemize}
		the analogy between this result and what was said in the Introduction is even stronger: after all, all these Lagrangian multipliers $ \lambda _{f} $, which are real numbers that cannot be unequivocally determined, are also class functions of the elements of $ G^{\prime } $ that lead us to an extrinsic parameterization of $ \mathcal{L} _{n} $.
		
		Of course, while these conclusions make sense, perhaps you, the reader, are seeing a difference that seems to be quite profound between the Hamiltonian formulation (\ref{lattice-hamiltonian-with-constraints}) and the one mentioned in the Introduction. After all, while the formulation mentioned in the Introduction points out that only the components of the pair $ \Omega _{I} = \left( Q_{I} , P_{I} \right) $ can be interpreted as gauge parameters, the formulation of Section \ref{constrained-characterisation} seems to be suggesting that both intrinsic and extrinsic parameters of the lattice gauge theories can be interpreted as such. And if you are seeing this difference, we need to remind you of an important mathematical result, which is directly related to the manifold parametrizations. And what important mathematical result is this? The one that tells us that, despite the parameters $ \omega $ are not necessarily identified as gauge parameters in (\ref{hamiltonian-reexpress}), the open sets of a differentiable manifold may be intrinsically parameterized in several ways [\refcite{man-gr,elon-variedades,man-gd}]. And since, for two open sets $ \mathcal{A} $ and $ \mathcal{B} $ such that $ \mathcal{A} \cap \mathcal{B} \neq \varnothing $, all these intrinsic parametrizations are related by diffeomorphisms [\refcite{man-gr,elon-variedades,man-gd}], this is another aspect that also ensures the covariance of the physical equations [\refcite{gitman,dirac,teitel}]. In other words, as
		\begin{itemize}
			\item all the faces of $ \mathcal{L} _{n} $ can be roughly interpreted as discretizations of the $ n $-dimensional open sets of $ \mathcal{M} _{n} $, and
			\item the intersection of two neighbouring faces of $ \mathcal{L} _{n} $ is always non-empty, since they always have at least a common edge,
		\end{itemize}	
		this situation is very similar to what is happening in (\ref{lattice-hamiltonian-with-constraints}) because both $ \omega $ and $ \psi \left( \mathfrak{g} _{\ell } \right) $ intrinsically parameterize the physical realities they describe.
		
		Anyway, note that, although most authors always present $ \mathcal{L} _{n} $ as a spacial lattice that is, at most, $ 3 $-dimensional, there is no obstacle for the partition function (\ref{pre-part-function}) to describe physical systems in spatial lattices with larger dimensions: the only conditions that $ \mathcal{L} _{n} $ needs to satisfy for this to happen is to
		\begin{itemize}
			\item support a gauge field on each of its edges, and
			\item (locally) discretize an orientable (sub)manifold $ \mathcal{M} _{n} $.
		\end{itemize}
		Surely, in view of our last comment, you, the reader, may be thinking that, when we deal with dimensionally larger lattices, other geometric information need to be invoked in order to better characterize this (sub)manifold that $ \mathcal{L} _{n} $ discretizes. And, if you are thinking about it, know that you are right and an excellent example of this is the $ \mathit{2} $\emph{-holonomies} [\refcite{2-gauge}] that appear in the lattice formulation of the \emph{higher gauge theories} [\refcite{baez}]. After all, while the usual lattice gauge theories are described by using only one gauge group, these \emph{higher lattice gauge theories} are defined by using two groups that, for instance, compose a crossed module [\refcite{white-1,white-2}]: one of these groups is the same $ G $ that defines the usual lattice gauge theories, while the other one describes other ($ 2 $-)holonomies that can be defined for the higher gauge fields [\refcite{2-gauge}]. In this fashion, it is clear that there is an open possibility to evaluate how these higher lattice gauge theories fit the description of a Hamiltonian system with constraints. As a matter of fact, it is worth noting that some generalizations of the $ D \left( G \right) $ models, which we cited as an example in the last Section, are being developed in the context of these higher lattice gauge theories [\refcite{2-gauge,pablo-1,pablo-2,pramod}]. %%%%%% 92A
		
	\section*{Acknowledgements}
	
		This work has been partially supported by CAPES (ProEx) and CNPq (grant 162117/2015-9). We thank R. A. Ferraz, A. F. Morais and P. Teotonio Sobrinho for some physical or/and mathematical discussions on subjects concerning this project. Special thanks are also due to the reviewer of this paper, who, although we do not know his/her/their name, made some relevant remarks that improved the text and made it more intelligible. %%%%%% Abstract and legends 90A

\end{document}